\documentclass[3p,times,twocolumn]{elsarticle}
\usepackage{ecrc1}
\usepackage{amsmath}

\volume{00}

\firstpage{1}

\journalname{Nuclear Physics B Proceedings Supplement}

\runauth{M.\ Czakon, M.\ Kr\"amer, and M.\ Worek}

\jid{nuphbp}

\jnltitlelogo{Nuclear Physics B Proceedings Supplement}

\usepackage{amssymb}

\biboptions{compress}

\usepackage[figuresright]{rotating}

\def\tb{\bar t}

\def\rel{\mathrm{rel}}
\def\th{\mathrm{th}}
\def\nlo{\mathrm{NLO}}
\def\nlops{\mathrm{NLO+PS}}

\usepackage{braket}
\usepackage{tikz}
\usepackage{tikz-cd}
\usetikzlibrary{matrix, calc, arrows}

\begin{document}

\begin{frontmatter}



\dochead{}

\title{Automated NLO/NLL Monte Carlo programs for the LHC}


\author[rwth]{Michael Czakon} \author[rwth,slac]{Michael Kr\"amer} 
\author[rwth]{Malgorzata Worek}

\address[rwth]{Institute for Theoretical Particle Physics and Cosmology, 
RWTH Aachen University, D-52056 Aachen, Germany}
\address[slac]{SLAC National Accelerator Laboratory, 
Stanford University, Stanford, CA 94025, USA}

\begin{abstract} The interpretation of experimental measurements at
the LHC requires accurate theoretical predictions for exclusive
observables, and in particular the summation of soft and collinear
radiation to all orders in perturbation theo\-ry.  We report on recent
progress towards the automated calculation of multi-parton LHC cross
sections at next-to-leading order in QCD, including the summation of
next-to-leading logarithmic corrections through the combination with
parton showers.
\end{abstract}


\end{frontmatter}

\section{Introduction}
\label{sec:introduction}

Theoretical calculations within fixed-order perturbation theory allow
for accurate predictions of inclusive observables like total cross
sections. The analysis and interpretation of experimental signatures
at the LHC, however, require theoretical predictions for exclusive
final states, i.e.\ predictions for differential distributions or
cross sections with cuts on kinematic variables.  Higher-order
calculations for such exclusive final states involve in general large
corrections from soft or collinear parton emission, which need to be
summed to all orders. An efficient way to achieve such a summation is
through a parton shower. Parton showers typically include the leading
logarithmic contributions from soft and collinear gluon or quark
emission to all orders in perturbation theory. They form a central
part of Monte Carlo event generators and are thus essential to connect
theoretical models with realistic experimental signatures.  On the
other hand, standard parton shower event generators often rely on
leading-order expressions for the hard scattering processes and can
therefore not predict inclusive cross sections accurately.

A central goal of recent and current theoretical work in LHC phenomenology is
thus the  automated calculation of next-to-leading (NLO) LHC cross
sections including the summation of large corrections from multiple
quark and gluon emission through parton showers. Such a calculation
should combine the accuracy of NLO predictions for inclusive cross
sections with the power of parton shower Monte Carlo programs to
reliably describe differential distributions and cross sections with
cuts on kinematic variables. A naive combination of NLO calculations
with parton showers would, however, lead to double counting of
higher-order contributions that are included in both the NLO cross
section and the parton shower.  These contributions have to be
identified and subtracted from the calculation by means of a matching
procedure. 

Much work in recent years has been devoted to the formulation of
matching schemes that allow for a consistent combination of parton
showers and NLO calculations including
loop-corrections~\cite{Dobbs:2001gb,Collins:2001fm, Chen:2001nf,
Frixione:2002ik, Kurihara:2002ne, Kramer:2003jk, Frixione:2003ei, Soper:2003ya,
Nason:2004rx, Nagy:2005aa, Bauer:2006mk, Giele:2007di, Frixione:2007vw, Bauer:2008qh,
Lavesson:2008ah, Hoeche:2011fd}.  So far, however, only few publicly
available computer codes exist that implement such schemes, including
in particular \textsc{Mc@NLO}~\cite{Frixione:2008ym}, 
\textsc{Powheg}~\cite{Alioli:2010xd},  \textsc{Sherpa}~\cite{Gleisberg:2008ta}
and \textsc{Madgraph5\_aMc@NLO}~\cite{Alwall:2014hca}.  The problem of
double counting contributions that are included in both the NLO
calculation and the parton shower has been solved in all approaches in
a similar way: The parton shower evolution is generically described by
a Sudakov factor of the form
\begin{displaymath}
\exp\left(-\int_{Q^2}^\infty \frac{d\bar q^2}{\bar q^2} \int_0^1 dz\
\frac{\alpha_s}{2\pi}\,P_{g/q}(\bar q^2,z)
\right)\,,
\end{displaymath}
shown here for gluon emission off a quark with virtuality $Q^2$. In
the collinear limit $\bar q^2 \to 0$ the function $P_{g/q}(\bar
q^2,z)$ reduces to the Altarelli-Parisi splitting function. The
Sudakov factor includes short-distance contributions which are also
part of the NLO cross section and which can be identified by expanding
the exponential in powers of $\alpha_s$. When combining parton showers
with NLO calculations these contributions need to be subtracted to
avoid double counting.


Early work by many authors has focused on implementing specific processes
into the \textsc{Mc@NLO} program and within the \textsc{Powheg}
framework.  While the original \textsc{Mc@NLO} code is very
successful and now includes a large number of processes, it is tied
strongly to the parton shower of \textsc{Herwig} (both \textsc{Herwig6} and
\textsc{Herwig++}). This restriction has been lifted recently within
\textsc{Madgraph5\_aMc@NLO}, which is not only fully automatic, but
also parton shower independent and may be used with various versions
of \textsc{Herwig} and \textsc{Pythia}. We note that \textsc{Sherpa}
contains an implementation of \textsc{Mc@NLO} as
well. \textsc{Powheg}, on the other hand, allowed for matching with
a generic parton shower from the beginning and could thus be
interfaced with various Monte Carlo programs. In practice, the
implementation of specific processes has been greatly simplified
thanks to the \textsc{Powheg-Box} tool. In each of the frameworks,
multiple parton emission is summed at the leading-logarithmic (LL)
level only (although leading-color subleading logarithmic effects
might be included in parton showers based on NLO subtraction schemes,
such as the Catani-Seymour subtraction), and the results can thus not
compete with the accuracy of dedicated resummation calculations which
are routinely performed at next-to-leading logarithmic accuracy (NLL).

The formulation and improvement of parton showers, which are in
general based on various assumptions and approximations, has also
been addressed by various authors recently~\cite{Gieseke:2003rz,
Sjostrand:2004ef, Nagy:2007ty, Dinsdale:2007mf, Schumann:2007mg,
Nagy:2008ns, Nagy:2008eq, Nagy:2009re, Skands:2009tb, Platzer:2009jq,
Nagy:2009vg, Nagy:2012bt, Larkoski:2013yi, Hartgring:2013jma,
Nagy:2014mqa, Nagy:2014nqa, Nagy:2014oqa}. Notably,
Refs.\,\cite{Nagy:2007ty, Nagy:2008ns, Nagy:2008eq,Nagy:2012bt} have
proposed a parton shower that includes quantum interference, spin
correlations and sub-leading color effects.  Interference is treated
in standard parton showers only approximately by means of angular
ordering, while spin information is generally ignored completely.
Spin correlations, on the other hand, are crucial for example to
explore new physics models in cascade decays at the LHC. While
Refs.\,\cite{Nagy:2007ty, Nagy:2008ns, Nagy:2008eq,Nagy:2012bt}
describe the theoretical formulation of parton showers with quantum
interference, the first implementation~\cite{Nagy:2014mqa} so far is
based  on the standard spin-averaged and leading-color treatment of
parton splitting only.

The set-up of the parton
shower also has implications for the matching with NLO calculations.
If the splitting functions of the parton shower, $P_{ij}(\bar q^2,z)$,
and the momentum mapping follow closely the definition of the
subtraction terms used to regularize the soft and collinear
divergences in the NLO calculation, then the matching is simplified
considerably. With that in mind, various authors have formulated
parton showers that are based on commonly used subtraction
schemes~\cite{Schumann:2007mg, Dinsdale:2007mf}.

To improve on existing NLO plus parton shower implementations, quantum
interference contributions, spin correlations and sub-leading color
effects in the parton shower should be included
systematically. Ultimately,  one would like to develop a fast and
numerically robust code for the automated calculation  of multi-parton
LHC cross sections at next-to-leading order, including the summation
of next-to-leading logarithmic corrections through combination with
parton showers.  In this contribution we shall describe recent
progress towards this goal. 

The article  is structured as follows. In
Section~\ref{sec:subtraction} we shall describe  the construction of a
NLO subtraction scheme as derived from a NLL parton shower, and its
implementation  into the program package \textsc{Helac-NLO}.
Section~\ref{sec:bbbb} reports on the application of such a scheme for
the automated calculation  of NLO-QCD corrections  to the production
of four bottom quarks at the LHC. Section~\ref{sec:nll_shower} finally
discusses  progress towards developing and implementing a NLL parton
shower and the matching with \textsc{Helac-NLO}. We conclude in Section~\ref{sec:conclusions}.

\section{Subtraction schemes for NLO-QCD calculations}
\label{sec:subtraction}

In the following, a new subtraction scheme based on a parton shower
introduced by Nagy and Soper~\cite{Nagy:2007ty} is described. This new subtraction  has
been implemented in the publicly available \textsc{Helac-Dipoles}
 software \cite{Bevilacqua:2013iha}  and has already been tested in
the calculation of the NLO QCD  corrections to   $pp\to
b\bar{b}b\bar{b} +X$ production at the LHC (see  
the next section for more details).
Our main motivation, however, was to provide a framework for a simple
matching between a fixed-order calculation and the new parton
shower. However, before addressing this issue, the problem of the
integration of subtraction terms over the unresolved phase space
needed to be solved. This is the non-trivial part of any subtraction
scheme. Contrary to the usual practice at NLO, we did not perform
involved analytic integrations, but rather used a semi-numerical
approach. After a suitable parameterization, numerical integrations,
inspired by recent NNLO methods \cite{Czakon:2010td}, have been
performed. This has allowed us to cover both massless and massive
cases with comparable effort. Let us emphasize that our semi-numerical
approach distinguishes our work from earlier publications on the
subject presented e.g.\  in
Refs.\,\cite{Chung:2010fx,Chung:2012rq,Robens:2013wga}.

Let us start with the inclusive NLO QCD cross section for a generic
process involving $m+1$ final-state QCD partons with  momenta $p_a+p_b
\to p_1+\cdots + p_{m+1}$, which can be written as follows
\begin{eqnarray*}
\sigma_{\rm NLO} & = & \int_{m} d\Phi_{m} \,\, \mathcal{A}^{B}(\{p\}_{m}) 
\,\, F_{m} \\ 
& + & \int_{m+1} d\Phi_{m+1} \,\, \mathcal{A}^{R}(\{p\}_{m+1}) \,\, F_{m+1} \\
& + &  \int_{m} d\Phi_{m} \,\, \mathcal{A}^{V}(\{p\}_{m}) \,\, F_{m}  \\
& + &  \int_0^1 dx \int_{m} d\Phi_{m}(x) \,\, \mathcal{A}^{C}(x,\{p\}_m) 
\,\, F_m
\end{eqnarray*}
where 
\begin{eqnarray*}
\mathcal{A}^{B} \equiv \vert \mathcal{M}^{\mbox{\scriptsize{Born}}} 
\vert^2 \,, ~~~~~~~\mathcal{A}^{R} \equiv \vert 
\mathcal{M}^{\mbox{\scriptsize{Real}}} \vert^2  \,,\;\;\; \:\: 
\end{eqnarray*}
\begin{eqnarray*}
\mathcal{A}^{V} \equiv 2\,\Re \left[ \mathcal{M}^{\mbox{\scriptsize{Born}}} 
\, ( \mathcal{M}^{\mbox{\scriptsize{1-Loop}}} )^* \right]\,,
\end{eqnarray*}  
and $\mathcal{M}^{\mbox{\scriptsize{Born}}}$,
$\mathcal{M}^{\mbox{\scriptsize{1-Loop}}}$,
$\mathcal{M}^{\mbox{\scriptsize{Real}}}$ describe the Born, one-loop
and  real-emission matrix elements, respectively. The  integration
measure for the $m-$ and  the $m+1-$parton  phase space is denoted by
$d\Phi_{m}$ and  $d\Phi_{m+1}$, whereas  $F_{m}$ and  $F_{m+1}$ are the
jet  functions.

For $m$ well separated  hard jets, the Born contribution is finite,
whereas the virtual and the  real-emission terms are individually
divergent due to the presence of  soft and collinear singularities.
For an infrared-safe definition of partonic jets, all soft and
collinear divergencies that affect the virtual and real corrections
should cancel  for the inclusive cross section, except for the
singularities arising from the  emission of nearly-collinear partons
off the initial state, which are absorbed into a re-definition of the
parton distribution functions (PDFs). This absorption is achieved by
introducing suitable  collinear counterterms, $\mathcal{A}^C$.
However, the individual pieces, $\mathcal{A}^{V}$ and
$\mathcal{A}^{R}$, still suffer from soft and collinear  divergencies
and cannot be integrated numerically in four dimensions.  To solve
this problem, local counterterms, $\mathcal{A}^D$, that are designed
to match the singular structure of the integrand in the soft and
collinear limits, can be introduced: 
\begin{equation*}
 \sigma_{\rm NLO}  = \int_{m} d\Phi_{m} \,\, \mathcal{A}^{B}(\{p\}_{m}) 
\,\, F_{m} 
\end{equation*}
\begin{equation*}
 +  \int_{m+1} d\Phi_{m+1} \, \left[ \mathcal{A}^{R}(\{p\}_{m+1}) 
\,\, F_{m+1} - \mathcal{A}^{D}(\{p\}_{m+1}) \,\,F_{m}  \right] 
\end{equation*}
\begin{equation*}
 +\int_0^1 dx \int_m d\Phi_{m}(x) \left[ \, \delta(1-x) 
\left( \mathcal{A}^V(\{p\}_m) + \int_1 \mathcal{A}^D(\{p\}_{m+1}) \right) 
\right.
\end{equation*}
\begin{equation*}
 \left. + \,\, \mathcal{A}^C(x,\{p\}_m) \, \right] \, F_{m} \,.  
\end{equation*}  
They  are defined on the $(m+1)$-parton phase space,
denoted $\{p\}_{m+1}$ and are subtracted from $\mathcal{A}^R$ and
added back to $\mathcal{A}^V$ after integration over the phase space
of the unresolved parton. This procedure, called subtraction method,
makes the integrals  individually   convergent and thus well suited
for a Monte Carlo integration. 

The construction of the local counterterms is inspired by the well
known  property of the universal factorization of QCD amplitudes in
the soft and  collinear limits.  The singular structure  of an 
$(m+1)$-parton squared amplitude for two partons $p_i$ and $p_j$  that
become collinear can be expressed as follows:
\begin{equation*}
\langle \mathcal{M}(\{p\}_{m+1}) \vert \mathcal{M}(\{p\}_{m+1}) 
\rangle _{\rm sing}
  \,  \approx \, 
\end{equation*}
\begin{equation*}
\langle \mathcal{M}(\{\overline{p}\}_{m}^{(ij)}) 
\vert \left( \mathbf{V}_{ij}^\dagger \cdot \mathbf{V}_{ij} \right) 
\vert  \mathcal{M}(\{\overline{p}\}_{m}^{(ij)}) \rangle  \;,
\end{equation*}
where $\vert\mathcal{M}(\{\overline{p}\}_{m})\rangle$ is an amplitude 
for $m$ on-shell external partons, $\mathbf{V}_{ij}$ is an operator 
acting on the spin part of the amplitude and $\{\overline{p}\}_{m}^{(ij)}$ 
describes the reduced $m$-parton kinematics in the 
limit where partons $p_i$ and $p_j$ become collinear.
The structure of the real-emission 
contribution can therefore be reduced to 
the product of a finite Born  amplitude squared
times a divergent, collinear splitting kernel $\mathcal{C}^{ij}$ associated 
with the splitting $\overline{p}_i \to p_i+p_j$ :
\begin{eqnarray*}
\mathcal{A}^{R}(\{p\}_{m+1}) \, \approx \,  
\mathcal{A}^{B}(\{\overline{p}\}_{m}^{(ij)}) \otimes 
\mathcal{C}^{(ij)}(\overline{p}_i;p_i,p_j) \;,
\end{eqnarray*}
where $\otimes$ denotes spin correlations.
When a parton $p_j$ becomes soft, on the other hand,  
factorization can be written in the following form:
\begin{equation*}
\langle \mathcal{M}(\{p\}_{m+1}) \vert \mathcal{M}(\{p\}_{m+1}) 
\rangle _{\rm sing}
  \,  \approx  
\end{equation*}
\begin{equation*}
\sum_{k \ne j} \langle \mathcal{M}
(\{\overline{p}\}_{m}^{(j)}) \vert \left( \mathbf{T}_{i} 
\cdot \mathbf{T}_{k} \right) \vert  
\mathcal{M}(\{\overline{p}\}_{m}^{(j)}) \rangle \;,
\end{equation*}
 where $\mathbf{T}_{i}$, $\mathbf{T}_{k}$ are operators
acting on the color part of the amplitude  and
$\{\overline{p}\}_{m}^{(j)}$ is the soft limit of the kinematical
configuration $\{p\}_{m+1}$.  In the limit where $p_j \to 0$, one can
write the real-emission  contribution as 
\begin{equation*}
\mathcal{A}^{R}(\{p\}_{m+1}) \, \approx \, \sum_{k \ne j} 
\mathcal{A}^{B}(\{\overline{p}\}_{m}^{(j)}) \otimes \mathcal{S}^{(kj)}
(\overline{p}_i,\overline{p}_k;p_i,p_k,p_j) \;,
\end{equation*}
where the factorization is expressed in terms of $m$ soft splitting kernels 
$\mathcal{S}^{(kj)}$, one for each external parton, and the symbol $\otimes$ 
denotes color correlations.

Factorization properties described so far dictate  general rules for
constructing  local counterterms  for a given subtraction method. In
the first step,  a complete set of transformations, which map the
original  $m+1$ partons phase space, $\{p\}_{m+1}$, into a new one,
$\{\tilde{p}\}_{m}$, that describes $m$ on-shell partons, needs to be
defined. Subsequently, a set of splitting functions
$\mathcal{D}^{(\ell)}(\{\tilde{p}\}_{m},\{p\}_{m+1})$, matching the
behavior of the soft and collinear kernels in the singular limits,
needs to be worked out. Consequently, the local counterterms take the
general form:
\begin{equation*}
\mathcal{A}^{D}(\{p\}_{m+1}) = \sum_{\ell=1}^{N} 
\mathcal{A}^B(\{\tilde{p}\}_{m}^{(\ell)}) \otimes 
\mathcal{D}^{(\ell)}(\{\tilde{p}\}_{m}^{(\ell)},\{p\}_{m+1}) \;.
\end{equation*}

There is a freedom in defining both mappings and 
splitting functions away from the singular limits, each choice leading 
to a different subtraction scheme. The most widespread version 
is the so-called Catani-Seymour scheme (CS) \cite{Catani:1996vz,Catani:2002hc}, where 
the local counterterms take the following form:
\begin{eqnarray*}
\lefteqn{\mathcal{A}^{D}_{\;\mbox{\scriptsize{CS}}}(\{p\}_{m+1}) = }\\
&&\sum_{i,j,k=1}^{m+1} \mathcal{A}^B(\{\tilde{p}\}_{m}^{(ijk)}) 
~\otimes~ \mathcal{D}^{(ijk)}_{\;\mbox{\scriptsize{CS}}}\;.
(\{\tilde{p}\}_{m}^{(ijk)},\{p\}_{m+1}) 
\end{eqnarray*}
Here, each mapping $\{\tilde{p}\}_{m}^{(ijk)}$ is labeled by 
three parton indices. For a large number, $m$, of external 
partons, the number of mappings and matrix elements required by the 
calculation scales cubically: $N_{\mbox{\scriptsize{CS}}} \sim m^3$. 
For our new Nagy-Soper subtraction scheme (NS), on the other hand, 
we can write
\begin{equation*}
\mathcal{A}^{D}_{\;\mbox{\scriptsize{NS}}}(\{p\}_{m+1})  = 
 \sum_{i,j,k=1}^{m+1} \mathcal{A}^B(\{\tilde{p}\}_{m}^{(ij)}) 
~\otimes~ \mathcal{D}^{(ijk)}_{\;\mbox{\scriptsize{NS}}}(\{\tilde{p}\}_{m}^{(ij)},
\{p\}_{m+1}) 
\end{equation*}
\begin{equation*}
=   \sum_{i,j} \mathcal{A}^B(\{\tilde{p}\}_{m}^{(ij)}) ~\otimes~ 
\left( \sum_{k} \mathcal{D}^{(ijk)}_{\;\mbox{\scriptsize{NS}}}
(\{\tilde{p}\}_{m}^{(ij)},\{p\}_{m+1})  \right) \;.
\end{equation*}
 Therefore, each mapping is characterized by two labels
$\{ij\}$ only, which implies that $N_{\mbox{\scriptsize{NS}}} \sim
m^2$, i.e.\  the number of mappings and subsequent matrix element
evaluations is reduced by a factor $m$ compared to the  
Catani-Seymour  case. 

Our particular  choice of the mapping and a form of splitting
functions for the Nagy-Soper subtraction scheme can be found 
in Ref.\,\cite{Bevilacqua:2013iha}. Let us only mention that the latter are
based on the original matrix elements  for $q\to qg$, $g\to q \bar{q}$
and $g\to g g $. Let us also add here that 
the general structure of the real-emission
contribution  to $\sigma_{\rm NLO}$ in this scheme takes 
the following form: 
\begin{equation*}
\sigma_{\rm RE}  \equiv  \int_{m+1} d\Phi_{m+1} \,\,
[ \,\, \mathcal{A}^{R}(\{p\}_{m+1}) \,\, F_{m+1} 
\end{equation*}
\begin{equation*} 
-  \sum_{i,k,j=1}^{m+1} \mathcal{A}^B(\{\tilde{p}\}_{m}^{(ij)}) \otimes
  \mathcal{D}^{(ijk)}(\{\tilde{p}\}_{m}^{(ij)},\{p\}_{m+1}) \,\, F_{m}
  \,\, ]  
\end{equation*}
\begin{equation*} 
+  \sum_{i,k=1}^{m} \, \int_m
d\Phi_{m} \,\, \mathcal{A}^B(\{p\}_m) \otimes
\mathbf{I}^{(ik)}(\epsilon,\{p\}_{m}) \,\, F_{m}  
\end{equation*}
\begin{equation*} 
+  \sum_{i=\{a,b\}} \, \sum_{k=1}^{m}\, \int_0^1 dx \int_m
d\Phi_{m}(x) \,\, \mathcal{A}^B(x,\{p\}_m)
\end{equation*}
\begin{equation*} 
 ~\otimes~ \left[
  \mathbf{K}^{(ik)}(x,\{p\}_{m})  +  \mathbf{P}^{(ik)}(x,\mu_F^2) \right] F_{m} \,,
\end{equation*}  
where  $\mathbf{I}(\epsilon)$ and $\mathbf{K}$/$\mathbf{P}$ correspond to the 
integrated subtraction terms. More precisely
$\mathbf{I}(\epsilon)$ encodes the full soft/collinear structure of
the matrix element in the form of single and double poles in $\epsilon
= (d-4)/2$, with $d$ the number of space-time dimensions, 
together with a finite part. The $\mathbf{K}$/$\mathbf{P}$ operator
consists of purely finite pieces coming from the initial-state
splitting ($\mathbf{K}$) as well as from the collinear counterterms
($\mathbf{P}$) and involves an additional integration over the
momentum fraction $x$ of an incoming parton after splitting.

An important feature of the Nagy-Soper mapping, that we would like to
emphasize, is a change of  all spectator momenta at the same
time. This fact impacts the way  the factorization of the phase space
is performed. While in the case of initial-state emission there are no
substantial differences with respect  to the Catani-Seymour scheme,
the factorization is derived in a slightly different way when the
splitting occurs in the final state.  The Lorentz invariant phase space 
for a generic final state with $m+1$ partons is organized in
terms of recursive splittings
\begin{equation*}
 d\Phi_{m+1}(p_i,p_j,k_1,\cdots,k_{m-1};Q) = 
\end{equation*}
\begin{equation*}
\frac{d^3p_i}{(2\pi)^d\,2p_i^0} \,  \frac{d^3p_j}{(2\pi)^d\,2p_j^0} \,  
\frac{d^3k_1}{(2\pi)^d\,2k_1^0}  \cdots  \frac{d^3k_{m-1}}{(2\pi)^d\,2k_{m-1}^0} ~\times
\end{equation*}
\begin{equation*}
(2\pi)^d \, \delta^d(Q-p_i-p_j-k_1-\cdots-k_{m-1})
\end{equation*}
\begin{equation*}
= \int_{K^2_{\rm min}}^{K^2_{\rm max}} \frac{dK^2}{2\pi} \,  \int_{P^2_{\rm min}}^{P^2_{\rm max}} 
\frac{dP^2_{ij}}{2\pi} \, d\Phi_{m-1}(k_1,\cdots,k_{m-1};K) ~\times~
\end{equation*}
\begin{equation*}
d\Phi_2(P_{ij},K;Q) \, d\Phi_2(p_i,p_j;P_{ij})  \;,
\end{equation*}
where 
\begin{equation*}
P_{ij}  =   p_i+p_j \,,  ~~~~~~ Q  =  K+P_{ij} \,,  
\end{equation*}
and K is the so-called collective spectator momentum that is built by
all  spectator momenta 
\begin{equation*}
K = \sum_{i=1}^{m-1} k_i   \,. 
\end{equation*}
Moreover 
\begin{eqnarray*}
K^2_{\rm min} & = & (m_{k_1}+\cdots+m_{k_{m-1}})^2  \\
K^2_{\rm max} & = &  (\sqrt{Q^2}-m_{p_i}-m_{p_j})^2 \\
P^2_{\rm min}  & = & (m_{p_i}+m_{p_j})^2 \\
P^2_{\rm max}  & = & (\sqrt{Q^2}-\sqrt{K^2})^2 \;.  
\end{eqnarray*}
The masses of the on-shell final-state  partons are denoted by
$m_{p_i},m_{p_j},m_{k_i}$,  while  $\sqrt{K^2}$ is the invariant mass
of the collective spectator. One can observe that
\begin{equation*}
 d\Phi_{m-1}(k_1,\cdots,k_{m-1};K) = d\Phi_{m-1}(\tilde{k}_1,\cdots,
\tilde{k}_{m-1};\tilde{K}) \;,
\end{equation*}
which follows from the fact that the mapping $K \to \tilde{K}$ 
is a Lorentz transformation, and the phase space is Lorentz invariant.
In the  frame where the total momentum $Q$ is at rest,
the two-body phase space can be parameterized in terms of angular 
variables,
\begin{equation*}
d\Phi_2(P_{ij},K;Q) = \frac{1}{8\,(2\pi)^{d-2}} \, 
\frac{\lambda(Q^2,P_{ij}^2,K^2)^{\frac{d-3}{2}}}{(Q^2)^{\frac{d-2}{2}}} \, 
\int d\Omega_{d-1} \;,
\end{equation*}
where $d\Omega_{d-1}$ represents the solid angle in $d$ dimensions and 
$\lambda$ is the standard K\"allen function,
\begin{equation*}
\lambda(x,y,z) = x^2 + y^2 + z^2 -2xy - 2xz - 2yz \;.
\end{equation*}
When the total 
momentum $Q$ is at rest, the integral $\int d\Omega_{d-1}$ for the two phase 
space elements $d\Phi_2(P_{ij},K;Q)$ and $d\Phi_2(\tilde{p}_{i},\tilde{K};Q)$ is the 
same. This implies that the Jacobian related to the mapping $P_{ij} \to 
\tilde{p}_i$ is given by
\begin{equation*}
d\Phi_2(P_{ij},K;Q) = \left( \frac{\lambda(Q^2,P^2_{ij},K^2)}
{\lambda(Q^2,m_i^2,K^2)} \right)^{\frac{d-3}{2}} \, 
d\Phi_2(\tilde{p}_i,\tilde{K};Q) \;.
\end{equation*}
The phase space for the final-state emission can  therefore be written 
in the fully factorized form 
\begin{equation*}
d\Phi_{m+1}(p_i,p_j,k_1,\cdots,k_{m-1};Q) = 
\end{equation*}
\begin{equation*}
d\Phi_m(\tilde{p}_i,\tilde{k}_1,\cdots,\tilde{k}_{m-1};Q) ~\times~ d\xi_{fin} \;,
\end{equation*}
where
\begin{equation*}
\label{eq:dxi_fin}
d\xi_{fin} = \frac{dP^2_{ij}}{2\pi} \, 
\left( \frac{\lambda(Q^2,P^2_{ij},K^2)}{\lambda(Q^2,m_i^2,K^2)}
\right)^{\frac{d-3}{2}} d\Phi_2(p_i,p_j;
\underbrace{\beta\,\tilde{p}_i+\gamma\,Q}_{P_{ij}})
\end{equation*}
is the measure of the splitting phase space in $d$ dimensions. The 
parameters $\beta$ and $\gamma$ are uniquely fixed by setting
\begin{equation*}
\tilde{Q} =  Q \;, ~~~~~~
\tilde{K}^2  =  K^2  \,, ~~~~~~
\tilde{p}_i^2 = m^2_i \,.
\end{equation*}
and are given by
\begin{eqnarray*}
\beta & = & 2 \, \sqrt{\frac{(P_{ij} \cdot Q)^2- P_{ij}^2 \, Q^2}{(m_i^2 + 2 \, 
P_{ij}\cdot Q-P_{ij}^2)^2 - 4 \, m^2_i \, Q^2}} \;,   \\
& & \nonumber \\
\gamma & = & \frac{2\,P_{ij}\cdot Q + \beta \, 
(P_{ij}^2 - 2 \, P_{ij}\cdot Q - m^2_i)}{2 \, Q^2} \;.
\end{eqnarray*}
Therefore, in the singular limit  one would simply have  $P_{ij}^2 =
m_i^2$, $\beta = 1$, $\gamma = 0$ and  $\tilde{p}_i = P_{ij}$.  
%
\begin{figure}[th!]
\begin{center}
\includegraphics[width=0.5\textwidth]{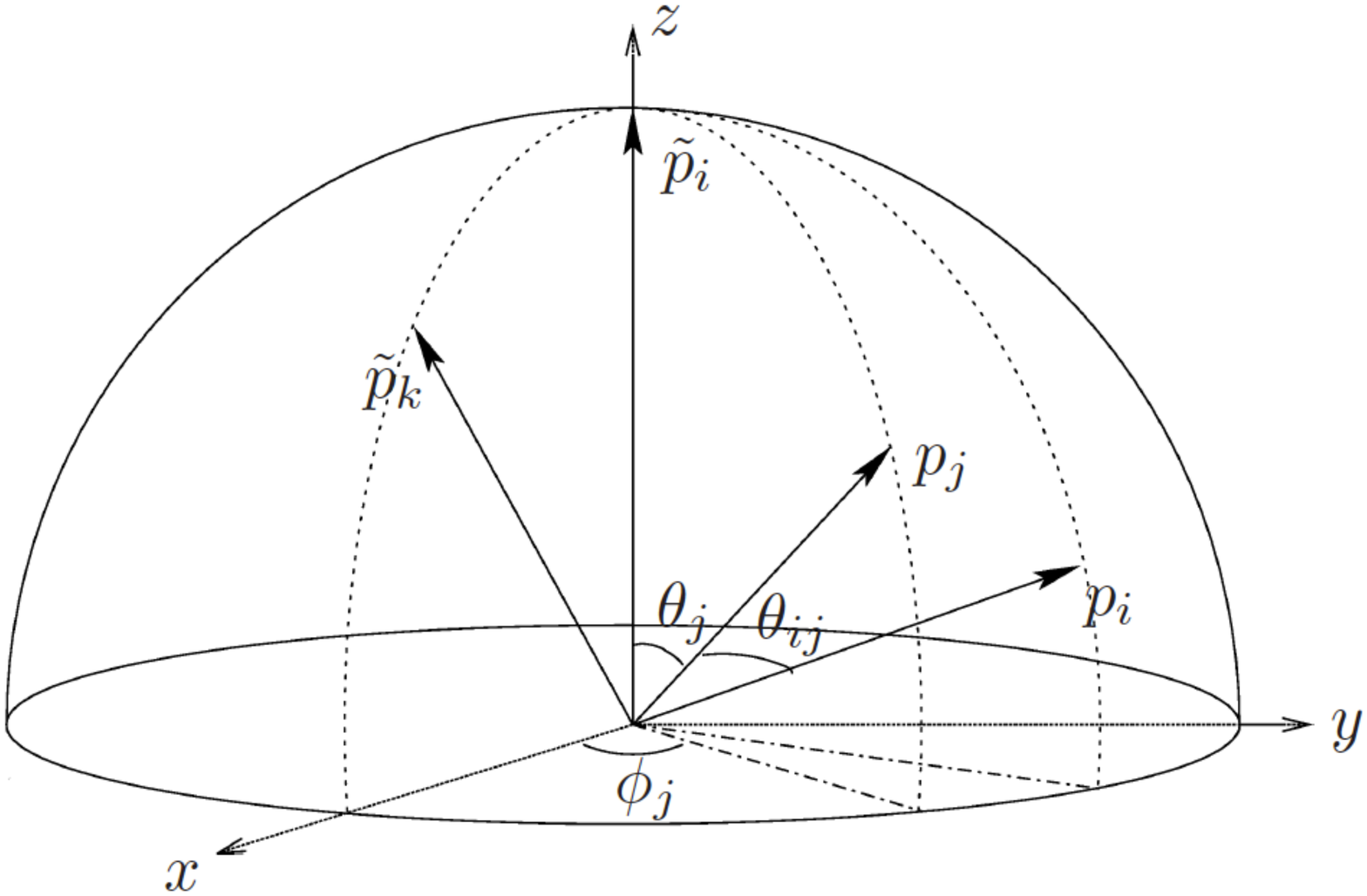}
\end{center}
\vspace{-1cm}
\caption{\it Parameterization of the angular variables for the final-state 
splitting $\tilde{p}_i \to p_i+p_j$. Here $\tilde{p}_k$ is the spectator 
parton selected to define the azimuthal variable $\phi_j$.}
\label{fig:angles_fin}
\end{figure}

The phase space of the splitting has three degrees  of freedom in
$d=4$ dimensions.  One possible way  of  parameterization is to use
Lorentz-covariant scalar pro\-ducts and  splitting variables  as
proposed  in Ref.\,\cite{Chung:2010fx}.  This simple choice   results
in compact formulae of the integrated dipoles for massless  partons
\cite{Chung:2012rq}. Applying the same strategy to the
fully massive case, however, the kinematical bounds of the splitting
become much more complicated and the resulting expressions turn out to be 
very cumbersome. To keep the final expressions
reasonably compact, we  have adopted an alternative  parameterization.
For example, in the case of  the final-state emission, for a set of
momenta $\{\tilde{p}\}_m$  the splitting  $\tilde{p}_i \to p_i+p_j$
and the set of momenta $\{p\}_{m+1}$ had to be constructed  out  of
three parameters that we have called collinear, soft and  azimuthal
variables.  In the singular limit, where $P_{ij}^2 \to m_i^2$,  the
collinear and the soft variables have to correspond to the relative
angle between the nearly-collinear partons,  $\theta_{ij}$, and  the
energy of the unresolved  parton, $E_j$.  On the other hand, the
azimuthal  variable,  $\phi_j$, which is the second angular para\-me\-ter
uniquely fixes  the  kinematics of the splitting.  Our reference frame
consists of  an orthogonal set of axes $(x,y,z)$, where the  $z$-axis
is identified with the spatial direction of the vector $\tilde{p}_i$,
as shown in  Figure\,\ref{fig:angles_fin}.  The azimuthal variable is 
then the angle,  which separates the 
unresolved parton $p_j$ from the the $x$-$z$ plane.  Since there is complete
 freedom in selecting the direction of the $x$-axis, we  have
decided to place it  in the plane $\tilde{p}_i$-$\tilde{p}_k$,  where
$\tilde{p}_k$ is the momentum of the spectator. In the frame, where the 
total momentum $Q$ is at rest, the definition of the soft variable is as 
follows:
\begin{equation*}
E_j \equiv \frac{\sqrt{Q^2}\,(P_{ij}^2-m_i^2)}{P_{ij}^2- m_i^2 + 
2\,\sqrt{Q^2}\,(\tilde{p}_i^0-\cos\theta_j\,\vert \vec{\tilde{p}}_i \vert)} \;.
\end{equation*}
It is convenient  to divide this soft variable by its 
kinematically allowed maximum value to get a normalized soft variable $e$:
\begin{equation*}
e \equiv E_j/E_{j}^{\rm max} \,.
\end{equation*}
The integration of the  splitting phase space in case of the final-state 
emission  runs over the variables 
\begin{equation*}
e  \in  [0,1] \;, 
\end{equation*}
\begin{equation*}
c \equiv \cos\theta_j  \in  [-1,1] \,,
\end{equation*}
\begin{equation*}
\phi \equiv \phi_j  \in  [0,2\pi] \;. 
\end{equation*}
The soft and collinear limits correspond to $e\to 0$ and $c\to 1$, 
respectively.  

For the calculation of the integrated subtraction terms in the
Nagy-Soper scheme, we have adopted the  spin-averaged version of the
splitting functions as described in  Ref.\,\cite{Nagy:2008ns}. Our goal is the  
integration in $d=4-2\epsilon$ dimensions over the whole phase
space of the splitting.  However, as a consequence of the increased
complexity of the mapping,  a fully analytic evaluation of the
integrals turns out to be demanding. Thus, as an  alternative we have used numerical
approaches to integrate over the splitting phase space. More
precisely, we have  decided to adopt a  semi-numerical approach to
consider analytic integration  when possible, and Monte Carlo
integration otherwise. Since  the general dependence of the integrands
on  the azimuthal variable $\phi$ was  simple and all the azimuthal
integrals could  be classified  into three groups, we carried out
this part of the integration analytically.  The dependence on the soft
and collinear variables was not as simple and  led to more complicated
expressions that we  treated numerically. Further details can be found
in the original publication  \cite{Bevilacqua:2013iha}. 

\subsection{Implementation in \textsc{Helac-Dipoles}}

We have incorporated the new subtraction method  based on the
Nagy-Soper formalism into the \textsc{Helac-Dipoles} package,
preserving at the same time all the  optimizations already available
in the code. For a detailed description of the  package
functionalities, we refer to the existing literature
\cite{Czakon:2009ss,Bevilacqua:2011xh}. All elements of the
calculation that  do not  dependent on a specific subtraction scheme,  like 
the Born matrix elements and the color correlators, were  
already provided by the framework of
\textsc{Helac-Dipoles}. This fact  has dramatically simplified our
implementation. 

The construction of the Nagy-Soper subtraction terms is dictated by
the  form of the splitting functions. They contain generic  spinors
and polarization  vectors, which enables them to treat simultaneously
fixed helicities as  well as  random polarization states.  We have
provided random polarization sampling as a further option  available
for the Nagy-Soper scheme. This is an alternative to the  existing
random helicity sampling optimization,  which uses stratified sampling
over the different (incoherent) helicity assignments  of partons
\cite{Czakon:2009ss}. The option for the spin sum treatment  can be
controlled by the user in the  configuration file
\texttt{dipoles.conf} as described in the Appendix of
Ref.\,\cite{Bevilacqua:2011xh}.

Besides random polarization sampling, which is an important speedup in
every calculation, we have also provided random sampling over color,
or color Monte Carlo, for the subtracted real radiation part. This
functionality has provided an important speedup for matrix elements
with a large number of colored external states. The general ideas from
\cite{Papadopoulos:2005ky,Bevilacqua:2009zn} have been adopted,
which  are also  an essential ingredient of  the 
\textsc{Helac-1Loop} package \cite{vanHameren:2009dr}. 

Our implementation of the Nagy-Soper subtraction scheme (NS) has been
tested and compared  to the Catani-Seymour subtraction scheme (CS)
for some specific processes.  More precisely  proton-proton collisions
at the LHC with a center-of-mass energy of 8 TeV have been considered
and the following partonic subprocesses $gg\to t\bar{t}b\bar{b}g$,
$gg\to t\bar{t}t\bar{t}g$, $gg\to b\bar{b}b\bar{b}g$, $gg\to
t\bar{t}ggg$ have  been studied.  They give dominant contributions to
the subtracted real emissions at ${\cal O}(\alpha_s^5)$  for the
corresponding processes  $pp \to t\bar{t}b\bar{b} +X$, $pp \to
t\bar{t}t\bar{t} +X$, $pp \to b\bar{b}b\bar{b} +X$ and $pp \to
t\bar{t}jj +X$. Moreover, they represent a high level of complexity and
test almost all aspects of the software, as they involve both massive
and massless states. We have imposed  basic selection cuts
on jets 
\begin{equation*}
 p_T(j)> 50 ~{\rm GeV}, ~~~~~~|y(j)|<2.5, ~~~~~~\Delta R(jj)>1\,,
\end{equation*}
which have been defined through the anti-$k_T$ jet algorithm \cite{Cacciari:2008gp} 
with radius parameter  $R = 1$. 
The mass of the top quark was set to $m_t = 173.5$ GeV and the
bottom quark was considered  to be massless.  Results have been presented for
the NLO CT10 parton distribution functions \cite{Lai:2010vv} with 
five active flavors
and the corresponding two-loop  $\alpha_s$.  The renormalization and
factorization scales were set to the scalar sum of the  jet transverse
masses
\begin{equation*} 
H_T=\sum_{i} m_T(j_i)\,, 
\end{equation*}  
where for the top quark 
\begin{equation*}
m_T(t)=\sqrt{m_t^2+p^2_T(t)}
\end{equation*} and for light
jets (also tagged bottom-jets) 
\begin{equation*}
m_T(j)=p_T(j) \,.
\end{equation*} 
A factor of $1/4$ has
been included in the scales 
for all but the $gg\to b\bar{b}b\bar{b}g$  process where
$\mu_R=\mu_F=H_T$ has been chosen instead. 

In the following, a few examples from  the comparison of both schemes are
given.   If not specified otherwise, full summation over all color
configurations has been  assumed together with random helicity sampling. 
%
\begin{table}[ht!]
\renewcommand{\arraystretch}{1.25} 
\setlength{\tabcolsep}{0.2pc}
\begin{center}
  \begin{tabular}{l|c|c|c}
 \textsc{Process}&
\textsc{Number of }   &
\textsc{Number of }  & \textsc{Number of  } \\
 \textsc{}&
\textsc{Dipoles (CS)}   &
\textsc{Dipoles (NS)}  & \textsc{FD} \\
\hline
$gg\to t\bar{t}b\bar{b}g$ & 55 & 11 & 341\\ 
$gg\to t\bar{t}t\bar{t}g$ & 30 & 6 &682 \\ 
$gg\to b\bar{b}b\bar{b}g$ & 90 &  18 &  682\\ 
$gg\to t\bar{t}ggg$ & 75 &  15 & 1240
  \end{tabular}
\end{center}
  \caption{\it \label{tab:real-emission-dipoles}  
Number of Catani-Seymour (CS) and Nagy-Soper (NS) subtraction terms 
for dominant partonic
subprocesses contributing to the subtracted real emission
contributions at ${\cal O}(\alpha_s^5)$ for the $pp \to
t\bar{t}b\bar{b} +X$, $pp \to 
t\bar{t}t\bar{t} +X$, $pp \to  b\bar{b}b\bar{b} +X$ and $pp \to
t\bar{t}jj +X$ processes at the LHC. The number of Feynman 
diagrams  (FD) corresponding to the subprocesses is given as well.}
 \end{table}
%
In  Table\,\ref{tab:real-emission-dipoles}, for example, the total
number of subtraction terms that are evaluated in both schemes is
shown. Also given is  the number of Feynman diagrams corresponding to
the subprocesses  under scrutiny to underline their complexity.
For each of the $2\to 5$ processes, five
times less terms are needed in the NS subtraction scheme compared to
the CS scheme.  The difference corresponds to  the total number of
possible spectators, which are relevant in the CS
case, but not in the NS case.
%
\begin{table}[ht!]
\renewcommand{\arraystretch}{1.25} 
\setlength{\tabcolsep}{0.1pc}
\begin{center}
  \begin{tabular}{l|c|c}
 \textsc{Process}&
 $\sigma_{\rm RE}^{\rm CS}$ [pb] &
$\sigma_{\rm RE}^{\rm NS}$ [pb] \\[0.5mm]
\hline
$gg\to t\bar{t}b\bar{b}g$ 
& $(28.39 \pm 0.04)\cdot10^{-3}$ 
& $(28.59 \pm 0.06)\cdot10^{-3}$ \\ 
$gg\to t\bar{t}t\bar{t}g$ 
& $(16.98 \pm 0.02)\cdot10^{-5}$ 
& $(17.01 \pm 0.03)\cdot10^{-5}$\\ 
$gg\to b\bar{b}b\bar{b}g$ 
& $(66.24 \pm 0.16)\cdot10^{-2}$ 
& $(66.06 \pm 0.22)\cdot10^{-2}$  \\ 
$gg\to t\bar{t}ggg$ 
& $(87.96 \pm 0.07)\cdot10^{-1}$
& $(88.16 \pm 0.08)\cdot10^{-1}$
  \end{tabular}
\end{center}
  \caption{\it \label{tab:real-emission-sigma}   Real emission cross
sections for dominant partonic subprocesses contributing to the
subtracted real emissions at ${\cal O}(\alpha_s^5)$ for the $pp \to
t\bar{t}b\bar{b} +X$, $pp \to t\bar{t}t\bar{t} +X$, $pp \to
b\bar{b}b\bar{b} +X$ and $pp \to t\bar{t}jj +X$ processes at the
LHC. Results are shown for two different subtraction schemes, the
Catani-Seymour (CS)  dipole subtraction 
and the new Nagy-Soper (NS) scheme, including the
numerical error from the Monte Carlo integration.}
 \end{table}
\begin{table}[ht!]
\renewcommand{\arraystretch}{1.25} 
\setlength{\tabcolsep}{0.5pc}
\begin{center}
  \begin{tabular}{l|c|c|c}
 \textsc{Process}&
$t^{\rm CS }$ [msec]
& $t^{\rm NS}$ [msec] &
$t^{\rm RE}$ [msec] \\
\hline
$gg\to t\bar{t}b\bar{b}g$ & $24.8$ & $13.2$ & $6.5$  \\ 
$gg\to t\bar{t}t\bar{t}g$ & $35.7$ & $18.5$ & $11.2$  \\
$gg\to b\bar{b}b\bar{b}g$ & $26.6$ & $16.2$ & $10.1$  \\
$gg\to t\bar{t}ggg$       & $214.8$ & $108.2$ & $48.7$  
  \end{tabular}
\end{center}
  \caption{\it \label{tab:subtracted-cpu}  The CPU time needed to
evaluate  the real emission matrix element together with all the
subtraction  terms for one phase space point for two
subtraction schemes, namely  Catani-Seymour, $t^{\rm CS }$, 
and Nagy-Soper, $t^{\rm NS}$.  
For comparison, we also give the CPU time
for the pure real emission matrix element calculation, $t^{\rm
RE}$. All numbers have  been obtained on an  Intel 3.40 GHz processor
with the Intel Fortran  compiler  using the option -fast.}
 \end{table}
\begin{table}[t!]
\renewcommand{\arraystretch}{1.25} 
\setlength{\tabcolsep}{0.1pc}
\begin{center}
  \begin{tabular}{l|c|c}
 \textsc{Process}&
 $\sigma_{\rm RE, \,COL}^{\rm CS}$ [pb] &
$\sigma_{\rm RE, \,COL}^{\rm NS}$ [pb] \\[0.5mm]
\hline
$gg\to t\bar{t}b\bar{b}g$ 
& $(28.35\pm 0.14)\cdot 10^{-3}$ 
& $(28.77\pm 0.14)\cdot 10^{-3}$  \\ 
$gg\to t\bar{t}t\bar{t}g$ 
& $(17.00\pm 0.03)\cdot 10^{-5}$ 
& $(17.01\pm 0.04)\cdot 10^{-5}$ 
\\ 
$gg\to b\bar{b}b\bar{b}g$ 
& $(65.71\pm 0.50)\cdot 10^{-2}$
& $(67.00\pm 0.66)\cdot 10^{-2}$
\\ 
$gg\to t\bar{t}ggg$ 
&  $(88.04\pm 0.37)\cdot 10^{-1}$
& $(87.76\pm 0.31)\cdot 10^{-1}$
  \end{tabular}
\end{center}
  \caption{\it \label{tab:real-emission-sigma-mc}   Real emission cross
sections for dominant partonic subprocesses contributing to the
subtracted real emissions at ${\cal O}(\alpha_s^5)$ for the $pp \to
t\bar{t}b\bar{b} +X$, $pp \to t\bar{t}t\bar{t} +X$, $pp \to
b\bar{b}b\bar{b} +X$ and $pp \to t\bar{t}jj +X$ processes at the
LHC. Results are shown for random color sampling for 
two different subtraction schemes, the
Catani-Seymour (CS)  dipole subtraction 
and the new Nagy-Soper (NS) scheme, including the
numerical error from the Monte Carlo integration.}
 \end{table}
\begin{table}[ht!]
\renewcommand{\arraystretch}{1.25} 
\setlength{\tabcolsep}{1.9pc}
\begin{center}
  \begin{tabular}{l|c}
 \textsc{Process}&
$\sigma_{\rm RE, \,POL}^{\rm NS}$ [pb]  \\[0.5mm]
\hline
$gg\to t\bar{t}b\bar{b}g$ 
& $(28.50\pm 0.06)\cdot 10^{-3}$ \\
$gg\to t\bar{t}t\bar{t}g$ 
& $(17.01 \pm 0.03)\cdot10^{-5}$ \\
$gg\to b\bar{b}b\bar{b}g$ 
& $(66.23 \pm 0.20)\cdot10^{-2}$  \\
$gg\to t\bar{t}ggg$ 
& $(88.16 \pm 0.07)\cdot10^{-1}$ 
  \end{tabular}
\end{center}
  \caption{\it \label{tab:real-emission-sigma-pol}   Real emission cross
sections for dominant partonic subprocesses contributing to the
subtracted real emissions at ${\cal O}(\alpha_s^5)$ for the $pp \to
t\bar{t}b\bar{b} +X$, $pp \to t\bar{t}t\bar{t} +X$, $pp \to
b\bar{b}b\bar{b} +X$ and $pp \to t\bar{t}jj +X$ processes at the
LHC. Results are shown for random polarization sampling  for 
the new Nagy-Soper (NS) subtraction scheme, including the
numerical error from the Monte Carlo integration.}
 \end{table}

Real
emission cross sections are presented in Table\,\ref{tab:real-emission-sigma}, 
again for the CS  dipole
subtraction  and the new NS scheme. All results have been  obtained
with the same Monte Carlo statistics and the resulting relative  errors are well
below $1\%$. We observe that  the  difference
between two evaluations of a given cross section is at most twice the  sum
of the corresponding errors. 
In Table\,\ref{tab:subtracted-cpu}, the time measured in
milliseconds, needed to evaluate  the real emission matrix element and
the subtraction terms for one phase space point is shown.
The NS is scheme is typically twice as fast as the CS scheme, but still a 
factor of about two slower than the evaluation of the real emission matrix element.   
Overall, both schemes, with their different momentum mappings and
subtraction terms, show a comparable performance and give the same  results
for total real emission  cross sections. 

The performance of Monte Carlo sampling over color and polarization
has also been studied.  In Table\,\ref{tab:real-emission-sigma-mc} we present 
real emission cross sections, which
have been evaluated  with random color sampling, for both 
subtraction schemes.  We observe agreement with the
results presented in Table\,\ref{tab:real-emission-sigma}, where
a summation over all color flows has been performed. One should note
that  in the case of the MC summation the absolute errors are $3-4$
times higher.  In order to obtain the same absolute errors as for results 
including a summation of color flows, $9-16$
times more events need to be evaluated. However, the average
number of color flows corresponding to a random color configuration,
which is evaluated per phase space point, is dramatically reduced.  The
overall time to obtain the same result is therefore substantially shortened.
Our  conclusion is thus that random color sampling is a powerful
approach, especially for processes where the number of gluons is higher and
exceeds the number of quarks.

Finally, in Table\,\ref{tab:real-emission-sigma-pol} real emission
cross sections for random polarization sampling  for  our new NS
subtraction scheme are shown.  They should be compared to the numbers
given in Table\,\ref{tab:real-emission-sigma}, where we have used  random
helicity sampling. Perfect agreement is found.

To conclude this section, a complete implementation of the Nagy-Soper
subtraction scheme both for massive and massless partons  is now
available in the \textsc{Helac-Dipoles} software.  By design, the
Nagy-Soper scheme has less kinematical mappings and is, therefore,
faster. On the other hand, we have observed that the absolute error of
the most costly (in terms of computational time) subtracted real
emission contribution was slightly worse for Nagy-Soper than for
Catani-Seymour. In the end, we conclude that both
schemes are similar in terms of efficiency. We did not consider differences
below a factor of two in error or time, which are moreover process
dependent, a reason to prefer either scheme.
There are two advantages of our implementation: First, we
can now perform better tests when calculating fixed order NLO  QCD
corrections  by computing real radiation in two different schemes. A
case study is described in the next section.  Second, the
integrated subtraction terms facilitate  the matching of the fixed order
calculation and the Nagy-Soper parton shower with quantum interference. 
This part is described in Section~\ref{sec:nll_shower}.

\section{A case study: NLO-QCD corrections to the production 
of four bottom quarks at the LHC}
\label{sec:bbbb}

The production of four bottom quarks, $pp \to b
\bar{b} b \bar{b} + X$, is an important background to various Higgs
analyses and new physics searches at the LHC, including for example
Higgs-boson pair production in two-Higgs doublet models at large
$\tan\beta$~\cite{Dai:1996rn}, or so-called hidden valley scenarios
where additional gauge bosons can decay into bottom
quarks~\cite{Strassler:2006im}. Accurate theoretical predictions for
the Standard Model production of multiple bottom quarks are thus
mandatory to exploit the potential of the LHC for new physics
searches. Furthermore, the calculation of the NLO
QCD corrections to $pp \to b \bar{b} b \bar{b} + X$ provides a
substantial technical challenge and requires the development of
efficient techniques, with a high degree of automation.  
In Ref.\,\cite{Bevilacqua:2013taa} we have
performed an NLO calculation of $ b \bar{b} b \bar{b} $ production at
the LHC with the \textsc{Helac-NLO} system~\cite{Bevilacqua:2011xh}. In
particular, we have presented results based on the Nagy-Soper subtraction scheme 
introduced in Section~\ref{sec:subtraction}. Two calculational schemes have been
employed, the so-called four-flavor scheme (4FS) with only gluons and
light-flavor quarks in the proton, where massive bottom quarks are
produced from gluon splitting at short distances, and the
five-flavor-scheme (5FS)~\cite{Barnett:1987jw} with massless bottom
quarks as partons in the proton. At all orders in perturbation theory,
the four- and five-flavor schemes are identical, but the way of
ordering the perturbative expansion is different, and at any finite
order the results do not match. Comparing the predictions of the two
schemes at NLO thus provides a way to assess the theoretical uncertainty
from unknown higher-order corrections, and to study the effect of the
bottom mass on the inclusive cross section and on differential
distributions. First NLO results for $pp \to b \bar{b} b \bar{b} + X$
in the 5FS have been presented in Ref.\,\cite{Greiner:2011mp}. 
In Ref.\,\cite{Bevilacqua:2013taa} we have not
only provided an independent calculation of this challenging process
with a different set of methods and tools, but also a systematic study
of the bottom quark mass effects by comparing the 5FS and 4FS
results. We note that NLO results for the production of four top quarks 
in hadron collisions have been discussed in Ref.\,\cite{Bevilacqua:2012em}.

The calculation of the process $pp \to b \bar{b} b \bar{b} + X$ at NLO
QCD comprises the parton processes $gg \to b \bar{b} b \bar{b}$ and
$q\bar{q} \to b \bar{b} b \bar{b}$ at tree-level and including
one-loop corrections, as well as the tree-level parton processes $gg
\to b \bar{b} b \bar{b}+g$,  $q\bar{q} \to b \bar{b} b \bar{b} + g$,
$gq \to b \bar{b} b \bar{b} + q$ and $g\bar{q} \to b \bar{b} b \bar{b}
+ \bar{q}$. In the four-flavor scheme $q \in \{u,d,c,s\}$, and the
bottom quark is treated massive. The bottom mass effects are in
general suppressed by powers of $m_{b}/\mu$, where $\mu$ is the hard
scale of the process, e.g.\ the transverse momentum of a
bottom-jet. Potentially large logarithmic corrections $\propto
\ln(m_b/\mu)$ could arise from nearly collinear splitting of
initial-state gluons into bottom quarks, $g \to b\bar{b}$, where the
bottom mass acts as a regulator of the collinear singularity. This
class of $\ln(m_{b}/\mu)$-terms can be summed to all orders in
perturbation theory by introducing bottom parton densities in the
five-flavor scheme. The 5FS is based on the approximation that the
bottom quarks from the gluon splitting are produced at small
transverse momentum. However, in our calculation we have required that all
four bottom quarks can be experimentally detected, and we have thus imposed
a lower cut on the bottom transverse momentum, $p_{T,b} \ge
p_{T,b}^{\rm min}$. As a result, up to NLO accuracy the potentially
large logarithms in the process $pp \to b \bar{b} b \bar{b} + X$ are
replaced by $\ln(m_{b}/\mu) \to \ln(p_{T, b}^{\rm min}/\mu)$, with
$m_{b} \ll p_{T, b}^{\rm min} \lesssim \mu$,  and are thus much less
significant numerically. Therefore, for the process at hand, the
differences between the 4FS and 5FS calculations with massive and
massless bottom quarks, respectively, should be moderate, but may not
be completely negligible. 

Our calculation has been performed with the automated \textsc{Helac-NLO}
framework~\cite{Bevilacqua:2011xh}, which includes
\textsc{Helac-1loop}~\cite{vanHameren:2009dr} for the evaluation of
the numerators of the loop integrals and the rational terms,
\textsc{CutTools}~\cite{Ossola:2007ax}, which implements the OPP
reduction method~\cite{Ossola:2006us,Ossola:2008xq,Mastrolia:2008jb,
Draggiotis:2009yb}  
to compute one-loop amplitudes,
and \textsc{OneLoop}~\cite{vanHameren:2010cp} for the evaluation of
the scalar integrals. The singularities for soft and collinear parton
emission are treated using subtraction schemes as implemented in
\textsc{Helac-Dipoles}~\cite{Czakon:2009ss}, see the discussion 
in Section~\ref{sec:subtraction}. The phase space
integration is performed with the help of the Monte Carlo generators
\textsc{Helac-Phegas}~\cite{Kanaki:2000ey,Papadopoulos:2000tt,Cafarella:2007pc}
and \textsc{Kaleu}~\cite{vanHameren:2010gg}, including
\textsc{Parni}~\cite{vanHameren:2007pt} for the importance sampling. 

The \textsc{Helac-Dipoles} package has been based on the standard
Catani-Seymour dipole subtraction formalism
\cite{Catani:1996vz,Catani:2002hc}. We have extended
\textsc{Helac-Dipoles} by implementing the new subtraction
scheme~\cite{Chung:2010fx,Chung:2012rq} based on the momentum mapping and
the splitting functions derived in the context of an improved parton
shower formulation by Nagy and Soper~\cite{Nagy:2007ty}, as described in 
Section~\ref{sec:subtraction}. The results presented in Ref.\,\cite{Bevilacqua:2013taa} 
have been the first 
application of the Nagy-Soper subtraction scheme for a $2 \to 4$
scattering process with massive and massless fermions. 

Below we shall present a number of selected numerical results for the 
$pp \to b \bar{b} b \bar{b} + X$ cross section at the LHC at the centre-of-mass
energy of ${\sqrt{s} = 14}$\,TeV. We discuss the impact of the NLO-QCD
corrections, and study the dependence of the results on the bottom
quark mass. 

Let us first specify the input parameters and the acceptance cuts we
impose. The top quark mass, which appears in the loop corrections, is
set to $m_{t} = 173.5$\,GeV~\cite{Beringer:1900zz}.  We combine
collinear final-state partons with pseudo-rapidity $|\eta| <5$ into
jets according to the anti-$k_T$ algorithm~\cite{Cacciari:2008gp} with
separation $R = 0.4$. The bottom-jets have to pass the transverse
momentum and rapidity cuts $p_{T,b} > 30$\,GeV and $|y_{b}|< 2.5$,
respectively.  The renormalisation and factorisation scales are set to
the scalar sum of the bottom-jet transverse masses, $\mu_{R} = \mu_{F} =
\mu_0 = H_{T}$, with 
\begin{equation*}
H_{T} =
m_{T,b}+m_{T,\bar{b}}+m_{T,b}+m_{T,\bar{b}}
\end{equation*}
and the transverse mass
\begin{equation*}
m_{T,b}=\sqrt{m^2_{b}+p^2_{T,b}} \,.
\end{equation*} 
For the five-flavor scheme
calculation with massless bottom quarks the transverse mass equals the
transverse momentum, $m_{T,b} = p_{T,b}$.  Note that the implementation 
of a dynamical scale requires a certain amount of care, as the subtraction 
terms for real radiation have to be evaluated with a different kinematical 
configuration specified by the momentum mapping of the subtraction scheme. 
Comparing the results as obtained with the 
Catani-Seymour subtraction and the Nagy-Soper scheme, which is based on 
a different momentum mapping, provides an important and highly non-trivial 
internal check of the calculation. 

\subsection{Massless bottom quarks within the five-flavor scheme}

The NLO predictions for the inclusive cross section are 
presented in Table\,\ref{tab:5fs4fs} for the NLO MSTW2008~\cite{Martin:2009iq} 
parton distribution function (pdf), 
with five active flavors and the corresponding two-loop $\alpha_{\rm
s}$. To study the impact of the higher-order corrections, we also show
leading-order results obtained using the MSTW2008  LO pdf sets and 
one-loop running for $\alpha_{\rm s}$. 

Varying the renormalisation and factorisation scales simultaneously about the
central scale by a factor of two, we find a residual scale uncertainty
of approximately 30\% at NLO, a reduction by about a factor of two
compared to LO. The $K$-factor, $K = \sigma_{\rm
NLO}/\sigma_{\rm LO} = 1.37$, is sizeable. Note, however, that the $K$-factor 
is an unphysical quantity and depends strongly on both the default choice of 
scale and the pdf set~\cite{Bevilacqua:2013taa}. 

\begin{figure}[t!]
\begin{center}
\hspace*{-5mm}\includegraphics[width=0.55\textwidth]{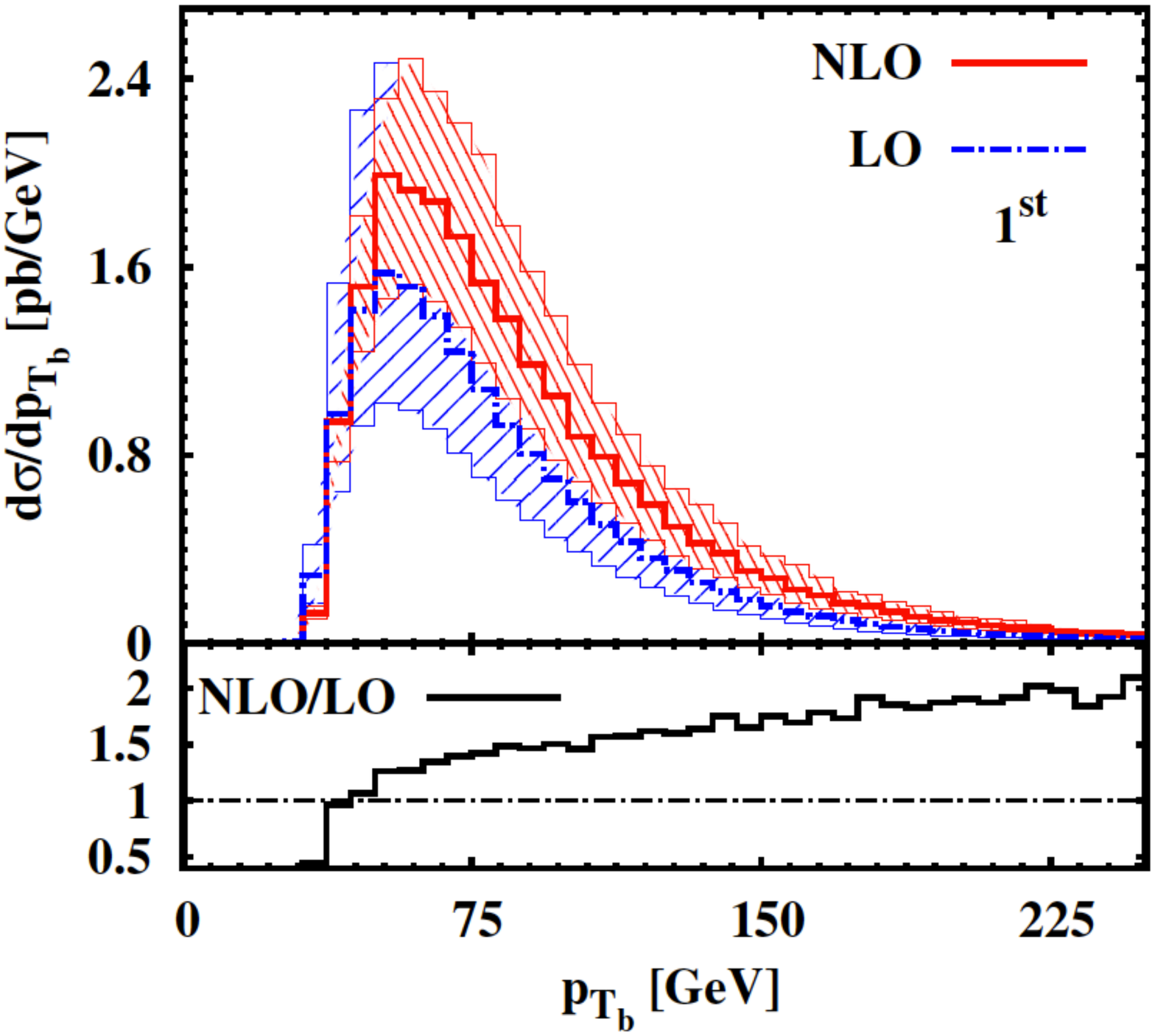}
\end{center}
\vspace*{-5mm}
\caption{\it \label{fig:nlo2:5f}    Differential cross section for
$pp\rightarrow b\bar{b} b\bar{b} ~+ X$ at the LHC ($\sqrt{s}$ = 14
TeV) in the 5FS as a function of the transverse momentum of the
hardest bottom jet. The
dash-dotted (blue) curve corresponds to the LO and  the solid (red)
curve to the NLO result. The scale choice is $\mu_R = \mu_F = \mu_0 =
H_T$. The hashed area represents the  scale uncertainty, and the lower
panels display the differential K factor. The cross sections  are
evaluated with the MSTW2008 pdf sets.}
\end{figure}

In Ref.\,\cite{Bevilacqua:2013taa} we have also  presented predictions for 
selected differential distributions which are an 
important input for the experimental analyses and the
interpretation of the experimental data. Figure\,\ref{fig:nlo2:5f} shows LO and 
NLO predictions for the transverse momentum of
the hardest bottom jet. We also show the theoretical
uncertainty through scale variation and the $K$-factor as a function
of the transverse momentum. It is evident from
Figure\,\ref{fig:nlo2:5f} that the NLO
corrections significantly reduce the theoretical uncertainty of the
differential distributions, and that the size of the higher-order
effects depends on the kinematics. For an accurate description of
exclusive observables and differential distributions it is thus not
sufficient to rescale the LO prediction with an inclusive $K$-factor.
%
\begin{center}
\begin{table*}[ht, width=\textwidth]
\renewcommand{\arraystretch}{1.25} 
\hfill{}
\begin{tabular}{c|c|c|c}
$pp\to b\bar{b}b\bar{b}+X$& $\sigma_{\rm LO}$\,[pb] & 
$\sigma_{\rm NLO}$\,[pb] & $K = \sigma_{\rm NLO}/\sigma_{\rm LO}$ \\[0.5mm] \hline 
5FS & $99.9^{+58.7\,(59\%)}_{-34.9\,(35\%)}$ 
& $136.7^{+38.8(28\%)}_{-30.9\,(23\%)}$ 
& 1.37\\
4FS & $84.5^{+49.7(59\%)}_{-29.6(35\%)} $ & 
$118.3^{+33.3(28\%)}_{-29.0(24\%)}$ & 1.40
\end{tabular}
\hfill{}
\caption{\it \label{tab:5fs4fs} 5FS and 4FS LO/NLO cross sections for
$pp\rightarrow b\bar{b} b\bar{b} ~+ X$ at the LHC ($\sqrt{s}$ = 14
TeV). The renormalisation and factorisation scales have been set to
the central value $\mu_0 = H_T$,  and the uncertainty is estimated by
varying both  scales
simultaneously by a factor two about the central scale.  Results are
shown for the 5FS/4FS MSTW2008LO/NLO pdf sets.} 
 \end{table*}
\end{center}

\subsection{Massive bottom quarks within the four-flavor scheme}
%
\begin{figure}[t!]
\begin{center}
\hspace*{-5mm}\includegraphics[width=0.55\textwidth]{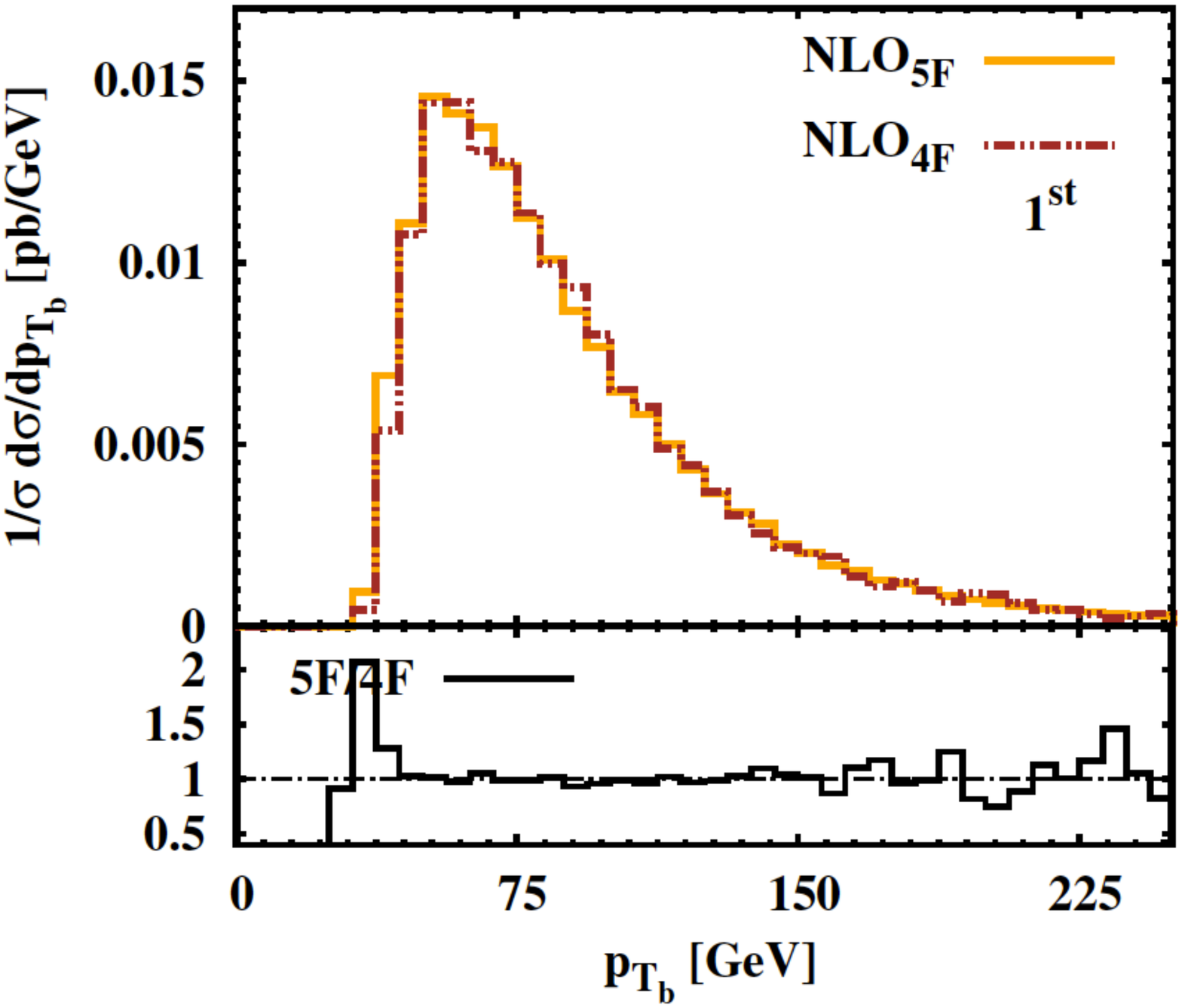}
\end{center}
\vspace*{-5mm}
\caption{\it \label{fig:nlo2:4f} Differential cross section for
  $pp\rightarrow b\bar{b} b\bar{b} ~+ X$ at the LHC ($\sqrt{s}$ = 14
  TeV) in the 4FS and 5FS  as a function of the transverse momentum of
  the hardest bottom jet, normalised to the corresponding inclusive cross section. 
  The scale choice is $\mu_R = \mu_F = \mu_0 = H_T$, the cross
  sections  are evaluated with the 5FS and 4FS MSTW2008 pdf sets,
  respectively.}
\end{figure}

Within the four-flavor scheme, bottom quarks are treated massive and
are not included in the parton distribution functions of the proton.
We define the bottom quark mass in the on-shell scheme and use 
$m_b = 4.75$\,GeV, consistent with the choice made in the  MSTW2008
four-flavor pdf~\cite{Martin:2010db}. 

The central cross section predictions in LO and NLO
for $\mu = H_T$ using the 4FS MSTW2008~\cite{Martin:2010db} pdf are
shown in Table\,\ref{tab:5fs4fs}, in comparison with the 5FS results. 
We observe that the bottom mass
effects decrease the cross section prediction by $18\%$ at LO and
$16\%$ at NLO. The residual scale dependence at NLO is approximately
30\%, similar to the 5FS calculation.

The difference between the massless 5FS and the massive 4FS
calculations has two origins.  First, there are genuine bottom mass
effects, the size of which depends  sensitively on the transverse
momentum  cut. For $p_{T,b}^{\rm min} = 30$\,GeV we find a 10\%
difference between the 5FS and 4FS  from non-singular bottom-mass
dependent terms. This difference decreases to about $1\%$ for
$p_{T,b}^{\rm min} = 100$\,GeV. Second, the two calculations involve
different pdf sets and  different corresponding $\alpha_{\rm
  s}$. While a 4FS pdf has, in general, a larger gluon flux than a 5FS pdf, as
there is no $g\to b \bar{b}$ splitting, the corresponding four-flavor
$\alpha_{\rm s}$  is smaller than for five active flavors. For
$pp\rightarrow b\bar{b} b\bar{b} ~+ X$ the difference in $\alpha_{\rm
  s}$ is  prevailing and results in a further reduction of the 4FS cross
section prediction by about $5\%$.  This latter difference
should be viewed as a scheme dependence rather than a bottom mass
effect.

In Figure\,\ref{fig:nlo2:4f} we present the differential distribution
in the transverse momentum of the hardest bottom jet, as calculated in
the 5FS with massless bottom quarks and in the 4FS with  $m_b =
4.75$\,GeV, normalised to the corresponding inclusive cross section.  
We find  that the difference in the shape of
the distributions in the 5FS and the 4FS is very small.

To conclude this section, we have presented selected results for
the differential cross-sections for $pp \to b \bar{b} b
\bar{b} + X$ at the LHC at the centre-of-mass energy of ${\sqrt{s} =
14}$\,TeV\,\cite{Bevilacqua:2013taa}. We find that the higher-order 
corrections significantly
reduce the scale dependence, with a residual theoretical uncertainty
of about 30\% at NLO. The impact of the bottom quark mass is moderate
for the cross section normalisation and negligible for the shape of
distributions. The fully differential NLO cross section calculation
for the process $pp\rightarrow b\bar{b} b\bar{b} ~+X$ 
presented in Ref.\,\cite{Bevilacqua:2013taa} 
provides an important input for the experimental analyses
and the interpretation of new physics searches at the LHC.

\section{Parton shower with quantum interference and matching}
\label{sec:nll_shower}

In this section, we will discuss the Nagy-Soper shower in more detail,
including its particular implementation in the C++ library
\textsc{Deductor} \cite{Nagy:2014mqa}.
Among other topics, we will elaborate on the theoretical framework
necessary to include quantum interference effects. Furthermore, we
will point out the inherent ambiguities of the approach. Ultimately,
we will discuss the matching to fixed order calculations at the
next-to-leading order in QCD, and present some results for a
non-trivial process: the production of a top-anti-top-quark pair in
association with a jet in hadronic collisions. This section is based
on \cite{in-preparation}.

\subsection{Basic concepts}

We start from a generic $2 \to  m$ process, which is defined by two
initial state partons $a$ and $b$ and $1,...,m$ final state
particles. Each particle is described by a set of quantum numbers to
define the flavor $f_i$, spin $s_i$ and color $c_i$ of the particle
and its momentum $p_i$. The initial state parton kinematics is described by
momentum fractions, $\eta_a$ and $\eta_b$, with respect to the original
colliding hadrons, instead of their momenta. Thus, a complete parton
ensemble can be described by\footnote{The minus sign for the initial
  state flavor is just a convention, because all partons are
  considered as outgoing.}
\begin{multline*}
 \{p,f,s,c\}_m \equiv \{ [\eta_a,-f_a, s_a,c_a], [\eta_b,-f_b,
 s_b,c_b],\\
[p_1,f_1, s_1,c_1],...,[p_m,f_m, s_m,c_m] \} \; .
\end{multline*}

The state of the parton shower evolution is described by a quantum
density matrix $\rho$, which gives the 'probability'\footnote{Since
  $\rho$ is related to color ordered amplitudes, it may become
  negative for subleading color configurations. Thus, one cannot
  naively interpret $\rho$ as a probability. Nevertheless, we use the
  terminology of statistical mechanics. } to find a
certain parton ensemble $\{p,f,s,c\}_m$. The
expectation value of an observable $F$ for any final state
multiplicity is given by
\begin{equation*}
\begin{split}
\sigma[F] =&
\sum_m \frac{1}{m!}\int [d\{p,f\}_m]
\frac{f_a(\eta_a,\mu_F^2)f_b(\eta_b,\mu_F^2) }{4n_c(a) n_c(b) \times
  \text{flux}} \\
&\quad \quad \quad \times \bra{M(\{p,f\}_m)}F(\{p,f\}_m)\ket{M(\{p,f\}_m)} \\
=& \sum_m \frac{1}{m!}\int [d\{p,f\}_m] Tr[ \rho(\{p,f\}_m)
F(\{p,f\}_m) ] \; .
\end{split}
\end{equation*}
Here the $1/m!$ is the symmetry factor for identical particles in the
final state and $[d\{p,f\}_m]$ is the $m$ particle phase space
measure. $f_{a/b}(\eta,\mu_F^2)$ are the parton density functions
evaluated at the momentum fraction $\eta$ and factorization scale
$\mu_F^2$. The factor $4$ in the denominator is the spin averaging
factor. $n_c(i)$ represents the averaging color factor for the initial
state partons, where $n_c(q) = 3$ and $n_c(g)=8$. The expectation
value of F and the trace in the second line are meant to be a
summation over indices in color $\otimes$ spin space. The quantum
density $\rho$ introduced in the second line, is thus given by
\begin{equation}
\begin{split}
\label{rho_def}
\rho(\{p,f\}_m) &= \;
\frac{f_a(\eta_a,\mu_F^2)f_b(\eta_b,\mu_F^2)}{4n_c(a) n_c(b) \times
  \text{flux}}  \\
&\times \ket{M(\{p,f\}_m)}\bra{M(\{p,f\}_m)} \\
&  \hspace*{-15mm}=\sum_{s,c} \sum_{s^\prime,c^\prime} \ket{\{s,c\}_m}
\rho(\{p,f,s^\prime,c^\prime,s,c\}_m) \bra{\{s^\prime,c^\prime\}_m} \; .
\end{split}
\end{equation}
It is useful to define basis vectors $|\{p,f,s^\prime,c^\prime,s,c\}_m)$,
such that
\begin{equation*}
\rho(\{p,f,s^\prime,c^\prime,s,c\}_m) =
(\{p,f,s^\prime,c^\prime,s,c\}_m|\rho) \; . 
\end{equation*}
One can then write the expectation value of the observable $F$ as
\begin{equation*}
 \sigma[F] = (F|\rho) \; .
\end{equation*}
A particularly important observable for the defintion of the parton
shower is the total cross section measurement function. It is defined
as
\begin{equation*}
 (1|\{p,f,s^\prime,c^\prime,s,c\}_m) = \braket{\{s^\prime\}_m|
   \{s\}_m} \braket{\{c^\prime\}_m| \{c\}_m} \; .
\end{equation*}

The shower evolution equation describes the propagation of the quantum
density matrix from some initial shower ``time'', $t_0$, which represents
the hard interaction, to the final ``time'' $t_F$ in the low energy
regime. The final shower time $t_F$ characterizes the physical scale
at which parton emissions cannot be described perturbatively
anymore. The definition of shower time $t$ is not unique and
is explained later. The parton shower evolution will transform a few
partons at the matrix element level, to a realistic final
state with jets typically made of many partons. After this evolution, a
phenomenological hadronization model must be applied. The 
perturbative evolution itself is described by a unitary operator
$U(t_F,t_0)$. The observable $F$, after showering, has the
expectation value
\begin{equation*}
\sigma[F] = (F|\rho(t_F)) = (F|U(t_F,t_0)|\rho(t_0)) \; .
\end{equation*}
The unitarity of the evolution operator is a consequence of the
requirement that it should not change the total cross section. Thus
$(1|U(t_F,t_0)|\rho(t_0))  = (1|\rho(t_0))$. The evolution operator
can be obtained from a differential equation involving two operators 
$\mathcal{H}_I(t)$ and $\mathcal{V}(t)$, corresponding to the concepts
of real and virtual corrections respectively:
\begin{equation}
  \frac{dU(t,t_0)}{dt} =[\mathcal{H}_I(t) - \mathcal{V}(t)]U(t,t_0) \; .
  \label{diffeq}
\end{equation}
Here  $\mathcal{H}_I(t)$ describes the emission of a resolved
particle, i.e.\ the momenta, flavor, spins and color configuration
will change after its application. $\mathcal{V}(t)$ describes the
unresolved emission and therefore does not alter momentum or flavor
configurations. Nevertheless it can change color configurations, which
will affect further emissions. For convenience of calculations, the
virtual operator can be further decomposed into $\mathcal{V}(t) =
\mathcal{V}_E(t) + \mathcal{V}_S(t)$, where $\mathcal{V}_E(t)$ is
diagonal in color space, while $\mathcal{V}_S(t)$ is
not. Interestingly, the evolution equation takes the same form as
the time evolution of a statistical ensemble in Liouville space
\begin{equation*}
 \frac{\partial\rho(t)}{\partial t} = \frac{i}{\hbar} [\rho(t),H] =
 L\rho(t) \; ,
\end{equation*}
where the Liouville operator can be identified as
$L =[\mathcal{H}_I(t) - \mathcal{V}(t)]$.

Traditional parton showers are constructed using the large $N_c$
limit. Thus, the state is always color diagonal implying
$\mathcal{V}_S(t) \to 0$. In this case, Eq.~\eqref{diffeq} yields
\begin{equation*}
 U(t,t_0) = N(t,t_0) + \int_{t_0}^t d\tau~
 U(t,\tau)\mathcal{H}_I(\tau)N(\tau,t_0) \; ,
\end{equation*}
with the Sudakov form factor
\begin{equation*}
 N(t,t_0) = \mathbb{T}\exp\left(-\int_{t_0}^t
   d\tau~\mathcal{V}(\tau)\right) \; .
\end{equation*}
Since $\mathcal{V}(t)$ is diagonal in the traditional approach,
$N(t,t_0)$ is a number and not a matrix in color space. In the general
case with non-diagonal $\mathcal{V}(t)$, it is not practical to
exponentiate a matrix in color space. The 
idea is, therefore, to exponentiate only the diagonal color part
$\mathcal{V}_E(t)$, and treat $\mathcal{V}_S(t)$ iteratively as a
perturbation. This can be justified by noting that the off diagonal
color contributions are always suppressed by a relative factor of
$1/N_c^2$ compared to the leading color contributions. Therefore,
the solution of Eq. \eqref{diffeq} with full color
evolution, using the decomposition of $\mathcal{V}(t) =
\mathcal{V}_E(t) + \mathcal{V}_S(t)$, is given by
\begin{multline*}
 U(t,t_0) = N(t,t_0) \\ + \int_{t_0}^t d\tau~
 U(t,\tau)\left[\mathcal{H}_I(\tau)-\mathcal{V}_S(\tau)\right]
 N(\tau,t_0) \; .
\end{multline*}
with
\begin{equation*}
 N(t,t_0) = \mathbb{T}\exp\left(-\int_{t_0}^t
   d\tau~\mathcal{V}_E(\tau)\right) \; .
\end{equation*}

\subsection{Real and virtual evolution operators}

The real evolution operator, $\mathcal{H}_I$, describes the transition
from an $m$-particle ensemble to an $(m+1)$-particle ensemble. This is
achieved by splitting a chosen parton into two, which would physically
correspond to a decay of a slightly off-shell parton. The splitting
is constrained by flavor conservation, $f_l \to \hat{f}_l +
\hat{f}_{m+1}$, and momentum conservation, $p_l \to \hat{p}_l +
\hat{p}_{m+1}$. The description of the transition is ambiguous, as
only the singular limits of amplitudes are uniquely
determined in QCD. After emitting a particle, it is necessary to
correct the momenta in the event in order to ensure momentum
conservation and preserve the on-shellness of all particles. This is
done by certain momentum mapping operators, which define momenta and
flavors of the new ensemble
\begin{equation}
\label{MomentumOperator}
	\{ \hat{p},\hat{f} \}_{m+1} = R_l(\{p,f\}_m) \; .
\end{equation}
where $l \in \{a,b,1,...,m\}$. In \textsc{Deductor}, a global momentum
mapping has been chosen. Whenever a particle is emitted, the momentum
of all final state particles is affected, see the discussion 
in Section~\ref{sec:subtraction}. In contrast, e.g.\ the
\textsc{Sherpa}~\cite{Gleisberg:2008ta} parton shower is based on
Catani-Seymour dipoles~\cite{Catani:1996vz}, which have local momentum
mappings. In this case a single parton momentum is modified. An
explicit description of the original momentum mapping used in
\textsc{Deductor} can be found either in Ref.\,\cite{Nagy:2007ty} or in
Ref.\,\cite{Bevilacqua:2013iha}. Recently, the initial state momentum
mapping has been slightly modified. A study~\cite{Nagy:2009vg} showed,
that the generated $p_T$ spectrum in $pp \to Z$ strongly depends on
the momentum mapping for initial state parton splittings. For this reason,
\textsc{Deductor} uses a momentum mapping, which allows for an
improved resummation of higher-order corrections~\cite{Nagy:2014nqa}.

Besides momentum and flavor mapping operators, $\mathcal{H}_I(t)$
contains splitting functions which correspond to the factorisation of
amplitudes in the soft, collinear and soft-collinear limits. In these
limits the amplitude can be written as
\begin{equation*}
	\ket{M(\{\hat{p},\hat{f}\}_{m+1})} =
        v(\{\hat{p},\hat{f}\}_{m+1}) \ket{M(\{p,f\}_m)} \; .
\end{equation*}
The operator $v(\{\hat{p},\hat{f}\}_{m+1})$ acts in color and spin
space. The behaviour of amplitudes in singular limits translates into
a similar behaviour of the density matrix, which can be written more
explicitly as
\begin{equation}
\begin{split}
\rho(\{\hat{p},&\hat{f}\}_{m+1}) \sim \\ &\sum_{l,k} T_l^\dagger(f_l
\to \hat{f}_l + \hat{f}_{m+1})
V_l^\dagger(\{\hat{p},\hat{f}\}_{m+1})\rho(\{p,f\}_m) \\ 
&\times V_k(\{\hat{p},\hat{f}\}_{m+1})T_k(f_k \to \hat{f}_k +
\hat{f}_{m+1}) \; ,
\label{rhoapprox}
\end{split}
\end{equation}
where $T_l^\dagger(f_l\to \hat{f}_l+\hat{f}_{m+1})$ is an operator in
color space, while $V_l^\dagger(\{\hat{p},\hat{f}\}_{m+1})$ is an
splitting operator in spin space. The general prescription to obtain
\textsc{Deductor}'s splitting functions $V_l$ has been summarised
in~\cite{Bevilacqua:2013iha}, whereas the complete set of splitting
functions can be found in~\cite{Nagy:2007ty}.

The approximation in Eq.~\eqref{rhoapprox} can be cast into an
operator equation, $|\rho_{m+1}) = \sum_l \mathcal{S}_l|\rho_m)$. The
operator $\mathcal{S}_l$ describes all possible splittings of the
emitter parton $l$. $\mathcal{H}_I(t)$ is then defined by the
splitting operators $\mathcal{S}_l$ at a fixed shower time
$\mathcal{T}_l(\{p,f \}_m$
\begin{equation}
\mathcal{H}_I(t) = \sum_l \mathcal{S}_l \delta\left( t
  -\mathcal{T}_l(\{p,f \}_m) \right) \; ,
\label{def:realsplittingoperator}
\end{equation}
where the sum runs over all possible emitters $l$. The shower time
$\mathcal{T}_l(\{p,f\}_m)$ corresponds to an infrared sensitive scale,
discussed in Section~\ref{subsec:time}.

The virtual evolution operator, $\mathcal{V}(t)$, represents the
unresolved virtual corrections. Nevertheless its content is fixed due
to the unitarity condition of the parton shower. By applying $(1|$
from the left and $|\rho)$ from the right to Eq. \eqref{diffeq} one
obtains
\begin{equation*}
(1|\mathcal{H}_I(t) - \mathcal{V}(t)|\rho(t)) = 0 \; ,
\end{equation*}
which should be valid for any $|\rho(t))$. Parton shower
unitarity corresponds to simplified virtual corrections, whose main
function is to cancel the divergences of the real
corrections. In order to obtain an expression for $\mathcal{V}(t)$, we
write
\begin{equation*}
(1|\mathcal{V}(t)|\{p,f,c^\prime,c,s^\prime,s\}_m) =
(1|\mathcal{H}_I(t)|\{p,f,c^\prime,c,s^\prime,s\}_m) \; .
\end{equation*}
This equation has an ambiguous solution in color space. We shall not
reproduce here the explicit form for $\mathcal{V}(t)$ used in
\textsc{Deductor}. It can be found in \cite{Nagy:2007ty}.

\subsection{Logarithmic accuracy}

In the previous section, we discussed the real and virtual evolution
operators.  While they allow for an exact treatment of color, the current
implementation of \textsc{Deductor} is based on the so-called LC+
approximation \cite{Nagy:2012bt}, which amounts to only allowing for
color evolution without non-local modifications of the color state. In
this case, we identify $\mathcal{V}_E(t) = \mathcal{V}^{\rm LC+}(t)$ and
$\mathcal{V}_S(t) =  \Delta\mathcal{V}(t) = \mathcal{V}(t) -
\mathcal{V}^{\rm LC+}(t)$.

The color diagonal part of the virtual operator will be exponentiated,
which yields a parton shower formulation similar to the traditional
one. The evolution equation is given by
\begin{multline*}
U^{\rm LC+}(t,t_0) = N^{\rm LC+}(t,t_0) + \\
\int_{t_0}^t d\tau~
U^{\rm LC+}(t,\tau)\mathcal{H}_I^{\rm LC+}(\tau)N^{\rm LC+}(\tau,t_0) \; ,
\end{multline*}
where
\begin{equation*}
  N^{\rm LC+}(t,t_0) = \exp\left( - \int_{t_0}^t
    d\tau~\mathcal{V}^{\rm LC+}(\tau)\right) \; .
\end{equation*}
The color off-diagonal splittings are included perturbatively
\begin{multline}
\label{fullEvo}
U(t,t_0) = U^{\rm LC+}(t,t_0) + \\
\int_{t_0}^t d\tau~ U(t,\tau) \left[\Delta\mathcal{H}_I(\tau) -
  \Delta\mathcal{V}(\tau)\right]U^{\rm LC+}(\tau,t_0) \; ,
\end{multline}
where $\Delta\mathcal{H}_I(t) = \mathcal{H}_I(t) -
\mathcal{H}_I^{\rm LC+}(t)$ and $\Delta\mathcal{V}(t) = \mathcal{V}(t) -
\mathcal{V}^{\rm LC+}(t)$. 

\begin{figure}[ht!]
\begin{center}
\includegraphics[width=0.5\textwidth]{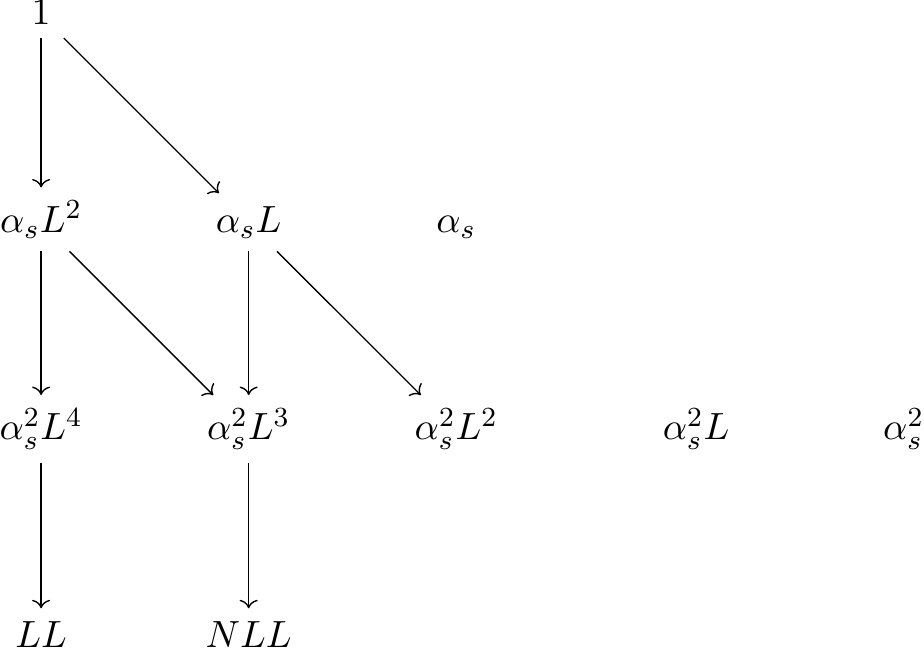}
\end{center}
\caption{Illustration of the logarithm counting. One step in the
  vertical direction is given by an insertion of $U^{\rm LC+}$ and a
  diagonal step is given by an insertion of $\Delta \mathcal{H}_I(t) -
  \Delta \mathcal{V}(t)$. One can see, that two insertions of $\Delta
  \mathcal{H}_I(t) - \Delta \mathcal{V}(t)$ only contribute to the
  coefficient of $\alpha_s^n L^{2n-2}$.}
  \label{fig:log_counting}
\end{figure}

Let us investigate the logarithmic accuracy of observables. We suppose
that the parton shower is used to calculate an observable
$\mathcal{O}$ which contains large logarithms L of some invariant. For
definiteness, we could image this invariant to be the transverse
momentum of a gauge boson. Then $\langle \mathcal{O} \rangle$ has the
form
\begin{align*}
\langle \mathcal{O} \rangle = \sum_n c(n,2n)\alpha_s^n L^{2n} 
+ \sum_n c(n,2n-1)\alpha_s^n L^{2n-1} + \cdots .
\end{align*}
We further suppose that a shower with full color, generated  by
$U(t,t_0)$ reproduces all coefficients $c(n,2n)$ and $c(n,2n-1)$
correctly. Then $U^{\rm LC+}(t,t_0)$ will reproduce $c(n,2n)$ exactly,
because the LC+ approximation is exact with respect to the
soft-collinear singularities. One insertion of
$\left[\Delta\mathcal{H}_I(\tau) - \Delta\mathcal{V}(\tau)\right]$
generates a contribution $\sim \alpha_s L$ because it contains a
correction for soft wide-angle gluon emission. This term
multiplies contributions of order $\alpha_s^{n-1}L^{2n-2}$ and,
therefore, corrects the coefficient $c(n,2n-1)$. A second insertion of
$\left[\Delta\mathcal{H}_I(\tau) - \Delta\mathcal{V}(\tau)\right]$
would only affect the coefficients $c(n,j \leq 2n-2)$. Therefore, one
insertion is sufficient to obtain NLL accuracy
\cite{Nagy:2012bt}.  (An illustration of the logarithm counting is given 
in Figure\,\ref{fig:log_counting}.) This implies that NLL accuracy is in fact
obtained with the evolution equation
\begin{multline*}
U(t,t_0) = U^{\rm LC+}(t,t_0) + \\
\int_{t_0}^t d\tau~ U^{\rm LC+}(t,\tau)
\left[\Delta\mathcal{H}_I(\tau) -
  \Delta\mathcal{V}(\tau)\right]U^{\rm LC+}(\tau,t_0) \; .
\end{multline*}

\subsection{Shower time}
\label{subsec:time}

Emissions generated by a parton shower are strongly ordered in
some kinematic variable in order to correctly resum the leading
logarithms. It turns out that the choice of the ordering variable,
which we call shower time, is ambiguous. The essential approximation
made in the parton shower description is that, in each step of the
evolution, all partons are on-shell. Thus, the parton shower time
should allow us to neglect the virtuality of a splitting parton. Here,
we summarize the discussion from Ref.\,\cite{Nagy:2014nqa} for final
state radiation. The discussion for initial state radiation is
analogous.

Consider a mother parton $0$ which splits in two partons, $p_0 \to p_1 +
p_2$. The momentum $p_0$ is given in light-cone\footnote{$k^\pm =
  (k^0\pm k^3)/\sqrt{2}$ and $k^2=2k^+k^- - \vec{k}^{~2}$} variables
$(+,-,\perp)$, where $\vec{p}_0$ denotes the transverse momentum. We
denote the virtuality of the parton as $v_0^2$. The momentum takes the
form
\begin{equation*}
	p_0 = \left(P,\frac{\vec{p}^{~2}_0 +
            m_0^2+v_0^2}{2P},\vec{p}_0 \right) \; .
\end{equation*}
The daughter momenta are given in the Sudakov parameterisation as
\begin{align*}
	p_1 &= \left(zP, \frac{\vec{p}^{~2}_1+m_1^2+v_1^2}{2zP},
          \vec{p}_1 \right) \; , \\
	p_2 &= \left((1-z)P,
          \frac{\vec{p}^{~2}_2+m_2^2+v_2^2}{2(1-z)P}, \vec{p}_2
        \right) \; .
\end{align*}
From momentum conservation  we obtain
\begin{equation*}
 v_0^2 = \frac{((1-z)\vec{p}_1-z\vec{p}_2)^2}{z(1-z)} +
 \frac{m_1^2}{z} + \frac{m_2^2}{1-z} -m_0^2 +\frac{v_1^2}{z}
 +\frac{v_2^2}{1-z} \; .
\end{equation*}
In order to be allowed to neglect $v_1^2$ and $v_2^2$ at each step of
the evolution, we must require
\begin{equation*}
  \frac{v_1^2}{z} \ll v_0^2 \; , \qquad \frac{v_2^2}{1-z} \ll v_0^2 \; .
\end{equation*}
Inserting the momentum fraction $z$
\begin{align*}
&z=\frac{p_1\cdot Q_0}{p_0\cdot Q_0} \; , \\
&1-z = \frac{(p_0-p_1)\cdot Q_0}{p_0\cdot Q_0} \approx \frac{p_2\cdot
  Q_0}{p_0\cdot Q_0} \; ,
\end{align*}
where the last approximation is valid in the singular limit $p_0
\approx p_1 + p_2$. Here, $Q_0$ is the total final state momentum. We
finally arrive at the conditions 
\begin{equation*}
  \frac{v_1^2}{2 p_1\cdot Q_0} \ll \frac{v_0^2}{2p_0\cdot Q_0} \quad
  \text{ and }\quad   \frac{v_2^2}{2 p_2\cdot Q_0} \ll
  \frac{v_0^2}{2p_0\cdot Q_0} \; .
\end{equation*}
They should always be fulfilled. Therefore, we define
\begin{equation*}
	\Lambda^2_l = \frac{|(\hat{p}_l \pm
          \hat{p}_{m+1})^2-m^2_l|}{2p_l\cdot Q_0} Q_0^2 \; ,
\end{equation*}
and enforce the emissions to be ordered in $\Lambda^2_l$.  This can be
achieved by defining the dimensionless shower time as
\begin{equation*}
 \mathcal{T}_l(\{p,f\}_m) = -
 \log\left(\frac{\Lambda_l^2}{Q_0^2}\right) \; .
\end{equation*}
Generally, one can choose other ordering variables. E.g.\
\textsc{Pythia 8} \cite{Sjostrand:2007gs} uses the transverse
momentum, $p_T$, in the parton splittings to order the emissions,
\textsc{Herwig} \cite{Bahr:2008pv,Corcella:2000bw} uses angualar ordering, and
\textsc{Pythia 6} \cite{Sjostrand:2006za} uses a virtuality ordering. A
consequence of the $\Lambda^2$ ordering is an enlarged phase space for
initial state splittings as compared to $p_T$
ordering~\cite{Nagy:2014nqa}.

\subsection{Ambiguities of the parton shower definition}

As we have already pointed out at various places, the construction of
the parton shower is not uniquely defined. In the following we list
the main ambiguities. They should be kept in mind, since future
findings might require modifications of their solutions.

\begin{description}

  \item[Momentum Mappings:] The way the parton shower distributes
    momentum among the particles after splitting may influence the
    resummation accuracy. In case of Drell-Yan Z-boson production, a
    study \cite{Nagy:2009vg} showed that a different momentum
    mapping \cite{Nagy:2014nqa} generates a different $p_T$ spectrum
    of the Z-Boson.

  \item[Splitting functions:] Splitting functions are only required to
    reproduce the singular limits of QCD matrix elements, but can have
    an arbitrary finite remainder.

  \item[Soft partition function:] Soft emissions are spread on
    different collinear emitters by means of a partition
    function. The latter is, however, arbitrary. In~\cite{Nagy:2008eq}
    it was argued, that the partition function should not depend on
    the emitter's energy. A choice, which has not been explored, would
    be to make the partition function spin dependent.

  \item[Color treatment:] In \textsc{Deductor}, the color diagonal part
    of the evolution is exponentiated, whereas the off-diagonal part
    is treated perturbatively. The separation depends on the
    representation of the color algebra, if the perturbative insertion
    of the off-diagonal color operator is
    truncated. In Ref.\,\cite{Platzer:2012np} a different 
    approach for a full color treatment in a parton shower has been
    proposed.

  \item[Spin treatment:] The spin basis (not discussed in this
    proceedings) is arbitrary. A different choice could modify the spin
    weights in the spin evolution of the shower.

  \item[Shower time:] The parton shower emissions are strongly
    ordered. As found in Ref.\,\cite{Nagy:2014nqa}, $\Lambda^2$
    ordering provides a wider phase space for initial state splittings
    than ordering in $p_T$. In many cases, nevertheless, different
    ordering variables give the same results. A counter example would
    be,  e.g.\ angular ordering vs. $p_T$ ordering. Angular ordering
    preserves color coherence of soft gluon emissions, whereas $p_T$
    ordering cannot account for this effect \cite{Abe:1994nj}.

  \item[PDF evolution:] Traditional parton showers interface
    LHAPDF~\cite{Whalley:2005nh} to obtain the ratio of
    PDFs occurring in the backward evolution for initial state
    radiation. Therefore, those parton showers explicitly depend on the
    PDF kernels they use. \textsc{Deductor} tries to minimize this
    dependence by evolving the PDFs according to the shower splitting
    functions. Since the splitting functions are not uniquely defined,
    the PDF evolution is also not unique. Additionally, if massive
    quarks are assumed in the initial state, then the mass dependent
    terms of the splitting kernel depend on the definition of the
    shower time.

\end{description}

\subsection{Matching at next-to-leading order}

Matching NLO calculations with parton showers is a widely explored
subject and there already exist several matching
schemes~\cite{Dobbs:2001gb,Collins:2001fm, Chen:2001nf,
Frixione:2002ik, Kurihara:2002ne, Kramer:2003jk, Frixione:2003ei, Soper:2003ya,
Nason:2004rx, Nagy:2005aa, Bauer:2006mk, Giele:2007di, Frixione:2007vw, Bauer:2008qh,
Lavesson:2008ah, Hoeche:2011fd}. The
most popular ones are the \textsc{Powheg}
method~\cite{Nason:2004rx,Frixione:2007vw} and the \textsc{Mc@NLO}
formalism~\cite{Frixione:2002ik,Frixione:2003ei}.  A general
comparison between those two major schemes can be found in
\cite{Hoeche:2011fd}. We choose to work in analogy to \textsc{Mc@NLO} instead
of \textsc{Powheg}, because we are looking for a general solution, which can be
easily automated. Before we discuss the problems of the \textsc{Mc@NLO}
formalism and their solutions, we want to give a brief
overview of the general objectives of parton shower matching. This
section presents original results obtained in \cite{in-preparation}.

Independently of the accuracy of the matching we obtain the following
benefits:
\begin{description}
  
\item[Connection to low energy physics:] Inclusive distributions are
  not affected by showering. Nevertheless, the evolution of partons
  down to a scale $t_F$ allows to include decays of unstable particles,
  as well as the non-perturbative hadronization and
  multiple interactions models.
  
\item[Logarithmic accuracy:] Infrared sensitive
  observables, which are ill defined at fixed order, are replaced by
  finite predictions due to resummation of large logarithms generated
  by collinear, soft and soft-collinear splittings.
  
\end{description}
Matching at the next-to-leading order gives us furthermore:
\begin{description}
  
\item[Cross section normalization at NLO:] When considering an
  inclusive observable $F$ we want to keep the fixed order
  normalization of the cross section. Thus, the parton shower must not
  modify the total cross section
  \begin{equation*}
    (F|U(t_F,t_0)|\rho(t_0)) = \sigma^{\rm NLO}[F] \; .
  \end{equation*}

\item[High-$\mathbf{p_T}$ emission according to matrix elements:] The parton
  shower is valid in the soft and collinear regime. Thus, a parton
  shower description of high $p_T$ emissions is not reliable. Since
  NLO calculations are indeed valid in this region, one wants to
  recover the NLO predictions for high $p_T$ emissions after
  showering.
  
\item[Meaningful events:] Matching to parton shower is the
  only way to define events at NLO. Without matching, the weights of
  the real matrix element and the subtraction terms belong to
  different kinematics and diverge separately. Due to the matching
  scheme, they are combined and one obtains real emission phase space
  configurations with a finite, but not necessarily positive, weight.

\end{description}

We start our discussion of matching from the quantum density
matrix. For a generic $2\to m$ process at NLO, one can write it in a
perturbative expansion
\begin{equation*}
	|\rho) = \underbrace{|\rho_m^{(0)})}_{\text{Born} ,
          \mathcal{O}(1)} +
        \underbrace{|\rho_m^{(1)})}_{\text{Virtual},
          \mathcal{O}(\alpha_s)} +
        \underbrace{|\rho_{m+1}^{(0)})}_{\text{Real},
          \mathcal{O}(\alpha_s)} + \mathcal{O}(\alpha_s^2) \; .
\end{equation*}
Note that we normalize the leading order contribution in the counting
of the coupling to be of order $1$. $|\rho_m^{(0)})$ and
$|\rho_{m+1}^{(0)})$ correspond to tree level matrix elements, whereas
$|\rho_m^{(1)})$ to the one-loop amplitude. The definitions of these
densities are analogous to the definition given in
Eq. \eqref{rho_def}. Based on this quantum density matrix, the
observable $F$ after showering naively reads
\begin{multline*}
\sigma [F]^{PS}  = (F|U(t_F,t_0)|\rho) = \\ \sum_{\lambda = m}^{\infty}
\frac{1}{\lambda!} \int [d\Phi_\lambda]
(F|\Phi_\lambda)(\Phi_\lambda|U(t_F,t_0)|\rho) \; ,
\end{multline*}
where we use the shorthand $\Phi_\lambda = \{p,f,s^\prime, c^\prime,
s,c\}_\lambda$. The quantum density $|\rho)$ accounts for the hard
matrix elements for $\lambda = m,m+1$. Finally $U(t_F,t_0)$ describes
the parton evolution during showering, as defined in
Eq. \eqref{fullEvo}.

This naive description of the cross section suffers from double
counting, which is demonstrated as follows. The evolution equation of
the parton shower has an iterative solution. Since, at first, we are only
interested in corrections up to $\mathcal{O}(\alpha_s)$, it is
sufficient to expand the evolution equation linearly. Evolving the
density state $|\rho)$  from $t_0$ to $t_F$ yields
\begin{multline*}
|\rho(t_F))= U(t_F,t_0)|\rho) \approx \\ |\rho) + \int_{t_0}^{t_F} d\tau
\left[ \mathcal{H}_I(\tau) - \mathcal{V}(\tau)\right]|\rho_m^{(0)})+
\mathcal{O}(\alpha_s^2) \; .
\end{multline*}
As we can see from the unitarity condition $(1|\left[
  \mathcal{H}_I(\tau) - \mathcal{V}(\tau)\right] = 0$, the total cross
section $(1|\rho(t_F))$ is conserved. On the other hand, for inclusive
observables $F$ one does not recover the NLO prediction, because in
general $(F|\left[ \mathcal{H}_I(\tau) - \mathcal{V}(\tau)\right] \neq
0$. Even if we would not insists on recovering the NLO prediction, the
result would have to be considered wrong, since it would contain real
emission contributions twice: from the real emission quantum 
density $|\rho^{(0)}_{m+1})$,  and from its parton shower
approximation $\mathcal{H}_I(\tau)|\rho^{(0)}_m)$.

This problem is solved by matching. The solution is slightly simpler
for processes, which can be defined without any cuts at the Born level,
e.g.\ $pp \to t\bar{t}$ or $pp \to W^+W^-$. At the end, we will
obtain a color and spin correlated version of the original
\textsc{Mc@NLO} formalism. The additional parton shower contribution
can be cancelled by including the same term with an opposite sign in
the quantum density state $|\rho)$. Thus, we can avoid double counting
by introducing a modified quantum density state
\begin{equation}
|\bar{\rho}) \equiv |\rho) - \int_{t_0}^{t_F} d\tau \left[
  \mathcal{H}_I(\tau) - \mathcal{V}(\tau)\right]|\rho_m^{(0)})+
\mathcal{O}(\alpha_s^2) \; .
\label{RhoBar}
\end{equation}
First notice that $(1|\bar{\rho}) = (1|\rho) = \sigma^{\rm NLO}$ is
unchanged, due to the unitarity condition. On the other hand,
considering $U(t_F,t_0)|\bar{\rho})$ and expanding the evolution
equation again shows that the undesired parton shower contributions
are cancelled up to $\mathcal{O}(\alpha_s)$. Notice that this
cancellation is non trivial, since the modified quantum density
$|\bar{\rho})$, depends now explicitly on the parton shower splitting
kernels and the choice of $t_0$.

Let us investigate the expectation value for an infrared safe
observable $F$ given by the density state in Eq. \eqref{RhoBar}
\begin{equation*}
\begin{split}
\bar{\sigma} [F] =&\frac{1}{m!}\int  [d\Phi_m] (F|U(t_F,t_0) |\Phi_m) \\
&\times \left[  (\Phi_m|\rho_m^{(0)})  + (\Phi_m|\rho_m^{(1)}) +
  \int_{t_0}^{t_F} d\tau (\Phi_m|V(\tau)|\rho_m^{(0)}) \right] \\ 
+&\frac{1}{(m+1)!} \int [d\Phi_{m+1}] (F|U(t_F,t_0) |\Phi_{m+1}) \\
&\times \left[ (\Phi_{m+1}|\rho_{m+1}^{(0)}) - \int_{t_0}^{t_F} d\tau
  (\Phi_{m+1}|H_I(\tau)|\rho_m^{(0)}) \right] \; .
\end{split}
\end{equation*}
Written in this way, the matched cross section suffers from infrared
divergences in the virtual $|\rho^{(1)}_m)$ and real contributions
$|\rho^{(0)}_{m+1})$, which appear in two separate integrals. In the
\textsc{Mc@NLO} approach one uses the parton shower splitting kernels
as subtraction terms, thus one drops the infrared cutoff, which is
imposed in the parton shower, and takes the limit $t_F \to \infty$. In
the subtracted real cross section we can make use of the definition of
the real splitting operator in Eq. \eqref{def:realsplittingoperator}
and write
\begin{multline*}
 \int_{t_0}^\infty d\tau~\mathcal{H}_I(\tau) = \\ \sum_l \mathbf{S}_l
 \int_0^\infty d\tau~\delta(\tau-t_l)\Theta(\tau - t_0) = \\ \sum_l
 \mathbf{S}_l \Theta(t_l - t_0) \; .
\end{multline*}
Here the sum runs over all external legs and $\mathbf{S}_l$ is the
total splitting kernel for a given external leg. We want to emphasize
that $\mathbf{S}_l$ also contains completely finite contributions like
the massive $g \to Q\bar{Q}$ splitting. $t_l$ is the shower time
defined in Section~\ref{subsec:time}. Hence $\Theta(t_l-t_0)$
represents the ordering of the emissions. The $t_0$ dependence
provides a dynamical restriction of the subtraction phase space. The
real subtracted cross section is now finite in $d=4$ dimensions, since
$t_l$ is allowed to approach infinity and therefore the subtractions
terms can resemble the singular limits of the QCD matrix element.

Integrating the virtual operator $\mathcal{V}(\tau)$ without an
infrared cutoff is more complex, since there is an explicit
integration over the splitting variables. Thus, we have to integrate
this part in $d=4-2\epsilon$ dimensions to extract the $1/\epsilon^2$
and $1/\epsilon$ poles analytically. $\mathcal{V}(\tau)$ takes the
form
\begin{multline*}
\int_{t_0}^\infty d\tau \mathcal{V}(\tau) = \sum_l \int d\Gamma_l\;
\mathbf{S}_l \Theta(t_l-t_0) = \\ \mathbf{I}(t_0) + \mathbf{K}(t_0) \; ,
\end{multline*}
where $\Gamma_l$ is the phase space of the additional parton.
The decomposition of 
the integrated $\mathcal{V}(\tau)$ into $\mathbf{I}(t_0)$,
which contains all integrated final-state splittings, and
$\mathbf{K}(t_0)$ which contains the initial-state splittings is
arbitrary. We emphasize this structure only to show that the parton
shower naturally incorporates a subtraction scheme as in the
Catani-Seymour framework \cite{Catani:1996vz}. The only part which
cannot be generated by the parton shower are the collinear
counterterms, denoted by $\mathbf{P}$, needed for the PDF
renormalization. Thus, the matched cross section reads
\begin{equation}
\begin{split}
\bar{\sigma} [F] &=\int \frac{[d\Phi_m]}{m!}  (F|U(t_F,t_0) |\Phi_m)
\\ \times& \left[ (\Phi_m|\rho_m^{(0)}) + (\Phi_m|\rho_{m}^{(1)})  +
  (\Phi_m|[\mathbf{I}(t_0) + \mathbf{K}(t_0) +
  \mathbf{P}]|\rho_m^{(0)} \right]  \\[0.2cm]
+\int & \frac{[d\Phi_{m+1}]}{(m+1)!}  (F|U(t_F,t_0) |\Phi_{m+1}) \\
\times& \left[ (\Phi_{m+1}|\rho_{m+1}^{(0)}) -  \sum_l
  (\Phi_{m+1}|\mathbf{S}_l|\rho_m^{(0)}) \Theta(t_l-t_0)\right] \; .
\label{Masterformula}
\end{split}
\end{equation}
In practice Eq.~\eqref{Masterformula} is not solved in a single
step. Let us define the shorthands (as in~\cite{Frixione:2002ik})
\begin{align*}
  (\Phi_m|S) \equiv \; & (\Phi_m|\rho_m^{(0)}) + (\Phi_m|\rho_{m}^{(1)})
  \nonumber \\ &+ (\Phi_m|[\mathbf{I}(t_0) + \mathbf{K}(t_0) +
  \mathbf{P}]|\rho_m^{(0)}) \; , \\[0.2cm]
  (\Phi_{m+1}|H) \equiv\; & (\Phi_{m+1}|\rho_{m+1}^{(0)}) -  \sum_l
  (\Phi_{m+1}|\mathbf{S}_l|\rho_m^{(0)}) \Theta(t_l-t_0) \; .
\end{align*}
One can use the total cross section
\begin{multline*}
\bar{\sigma}^{NLO} [1] =\frac{1}{m!}\int [d\Phi_m]  (1|\Phi_m)
(\Phi_m|S) \\ + \frac{1}{(m+1)!} \int [d\Phi_{m+1}]  (1|\Phi_{m+1})
(\Phi_{m+1}|H) \; ,
\end{multline*}
to generate the hard events according to $(\Phi_m|S)$ and
$(\Phi_{m+1}|H)$. The generated events are subsequently interfaced to
the parton shower and after the evolution one can investigate the
desired observables. Thus, after showering one has performed the
following integrals
\begin{multline*}
\bar{\sigma} [F]^{PS} =\frac{1}{m!}\int [d\Phi_m]
(F|U(t_F,t_0)|\Phi_m) (\Phi_m|S)  \\ + \frac{1}{(m+1)!} \int
[d\Phi_{m+1}]  (F|U(t_F,t_0)|\Phi_{m+1}) (\Phi_{m+1}|H) \; .
\end{multline*}

In the case of processes, which require cuts already at Born level to
yield a finite cross section, one has to modify the matching
prescription Eq.~\eqref{Masterformula}. Naively, one would simply
introduce a generation cut function given by a state $(F_I|$ as follows
\begin{align*}
 (\Phi_m|S) &\to (\Phi_m|S)(F_I|\Phi_m) \; , \\
 (\Phi_{m+1}|H) &\to (\Phi_{m+1}|H)(F_I|\Phi_{m+1}) \; .
\end{align*}
Applying the parton shower to these ensembles, shows that double
counting is still present \cite{in-preparation}. It turns out that it is also
necessary to modify $(\Phi_{m+1}|H)$ to be defined with account of the
generation cut in the subtraction phase space
\begin{multline*}
(\Phi_{m+1}|H) \to (\Phi_{m+1}|\tilde{H}) \equiv
(\Phi_{m+1}|\rho_{m+1}^{(0)}) \\ -  \sum_l
(\Phi_{m+1}|\mathbf{S}_l|\rho_m^{(0)}) (F_I|Q_l|\Phi_{m+1}) 
\Theta(t_l-t_0) \; ,
\end{multline*}
where we have introduced the momentum mapping operator $Q_l$ with
\begin{equation*}
	Q_l|\Phi_{m+1}) = |\Phi_m(\{\hat{p},\hat{f}\}_{m+1})) \; .
\end{equation*}
as the inverse transformation to that given by the operator $R_l$ in
Eq. \eqref{MomentumOperator}. It can be shown that this matching
prescription is correct as long as the generation cuts are looser than
those implied by the final observable, and the latter also only
amounts to cuts.

\subsection{Application: $t\tb j$ production at the LHC}

The matching scheme of the previous subsection has been implemented in
the framework of \textsc{Helac-NLO}. In order to test the
implementation, we have chosen to study the process of top-quark
pair production in association with an additional jet at the Large
Hadron Collider. The results reported in this subsection are taken
from \cite{in-preparation}. We point out that the NLO QCD corrections to $t\tb j$
production have been previously obtained in
Refs.\,\cite{Dittmaier:2007wz,Dittmaier:2008uj,Melnikov:2010iu,Melnikov:2011qx}.
Furthermore, NLO + parton shower predictions have been studied in
Ref.\,\cite{Alioli:2011as,Kardos:2011qa}.

Results for $t\tb j$ production are given for $pp$ collisions at the
LHC with a center-of-mass energy of 8 TeV. The top quark is assumed to
be stable and its mass is set to $m_t = 173.5$ GeV, while the bottom
quark is considered as massless. We use the NLO MSTW2008 PDF
set~\cite{Martin:2009iq} with five active flavors and the
corresponding two-loop running of the strong coupling. We set the
renormalization and factorization scale to the top quark mass, $ \mu_R
= \mu_F = \mu = m_t $, and the starting shower time $t_0$ to $T_0$, with
\begin{equation*}
e^{-T_0} = \min_{i\neq j}\left\{\frac{2 p_i \cdot p_j}{Q_0^2}\right\}
\; ,
\end{equation*}
where $Q_0^2$ is the partonic center-of-mass energy. Since the $t\tb
j$ process is divergent already at leading order, we have to impose
cuts on the hard jet in the event generation. These cuts have to be as
minimal as possible to ensure the inclusiveness of the events before
they are passed to the parton shower. We require the reconstructed
jets to have
\begin{equation*}
 p_T(j)  > 10\text{~ GeV} \;,  ~~~~~~~
 |y(j)| < 5 \;,
\end{equation*}
in the event generation, and 
\begin{equation*}
 p_T(j)  > 50\text{~ GeV} \;,  ~~~~~~~
 |y(j)| < 5 \;,
\end{equation*}
in the final analysis of several observables. Jets are clustered using
the anti-$k_T$ jet algorithm~\cite{Cacciari:2008gp}, with $R=1$
used at both the generation and analysis levels. Only particles with
pseudo-rapidity $|\eta| < 5$ are passed to the jet algorithm. In the
parton shower, the top quark is kept as a stable particle (i.e.\ no
decay allowed), hadronization and multiple interactions are not
included. In order to address the theory uncertainties we investigated
the scale dependence on the unphysical scales $\mu$ and $T_0$. Here
$\mu$ is varied between $\mu = m_t/2$ and $\mu = 2m_t$. Whereas
the parton shower starting time $t_0$ is varied between $t_0 =
T_0/\sqrt{2}$ and $t_0 = T_0\sqrt{2}$.

First we check the total cross section for $t\tb j$
production. Including scale variation of $\mu_R$ we obtain our final
prediction as
\begin{align*}
	&\sigma^{\nlo}(pp \to t\tb j + X) =
        86.04^{+5.10~(+6\%)}_{-11.41~ (-13\%)} \text{~pb} \; , \\
    &\sigma^{\nlops}(pp \to t\tb j + X) =
    85.94^{+3.81~(+4\%)}_{-11.43~ (-13\%)} \text{~pb} \; .
\end{align*}
These results are fully consistent, which proves that the initial cuts
during the generation phase were chosen appropriately. Additionally we
observe that the parton shower does not improve the theory uncertainty
of the total cross section at all. This is expected because we removed
all parton shower effects for the total cross section by
construction. Nevertheless, we expect some reduction of the scale
dependence for differential distributions, because of the summation of
the leading logarithms. 

\begin{figure}[t!]
\begin{center}
\includegraphics[width=7cm]{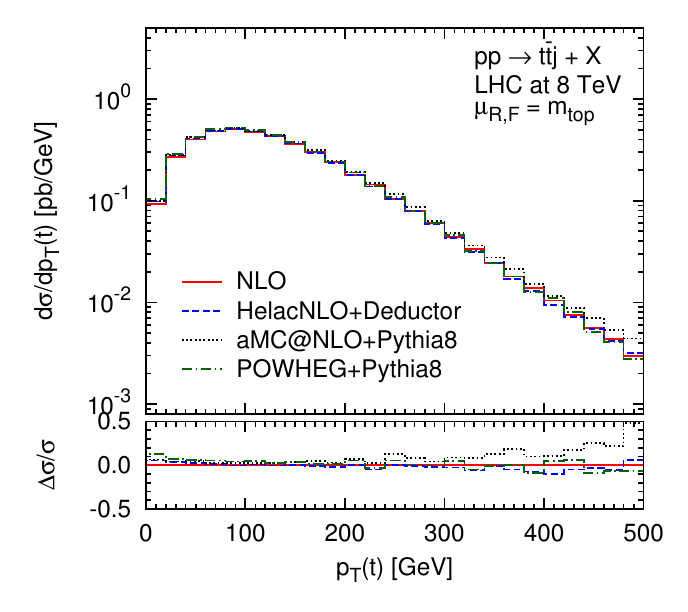} \\
\includegraphics[width=7cm]{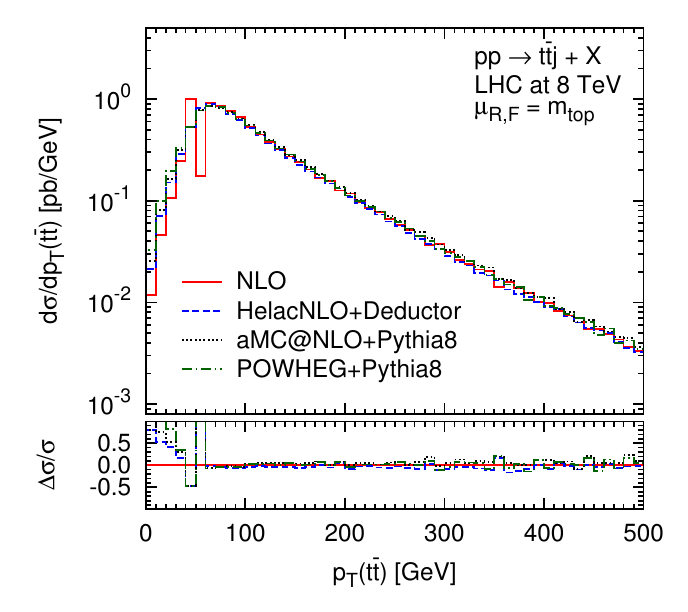} \\
\includegraphics[width=7cm]{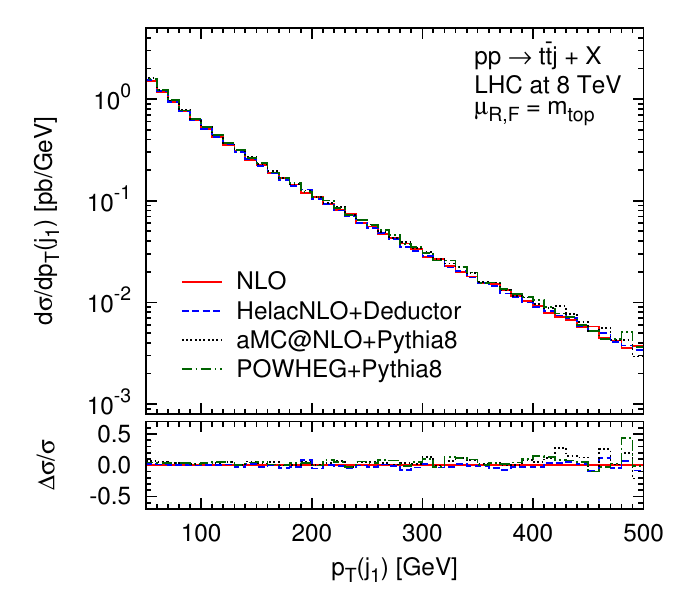}
\end{center}
\caption{Differential distributions for $t\bar{t}j$
  production. Comparison between NLO and several NLO+PS predictions.}
\label{fig:dist1}
\end{figure}

\begin{figure}[t!]
\begin{center}
\includegraphics[width=7cm]{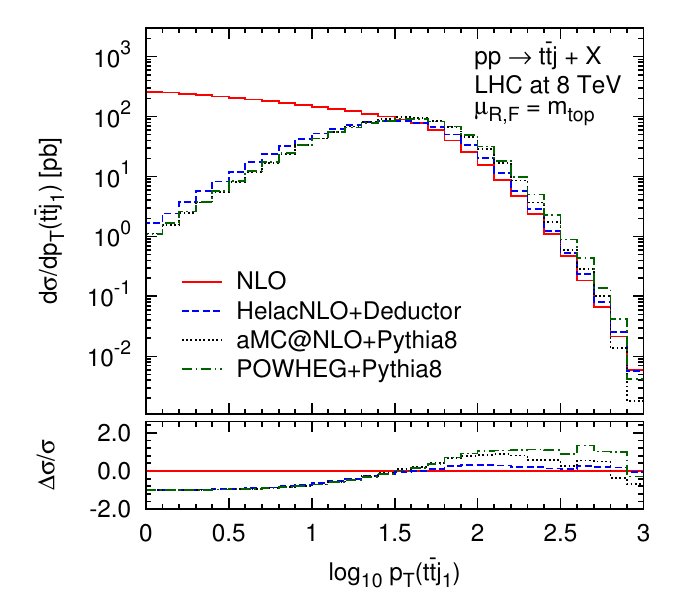}\\
\includegraphics[width=7cm]{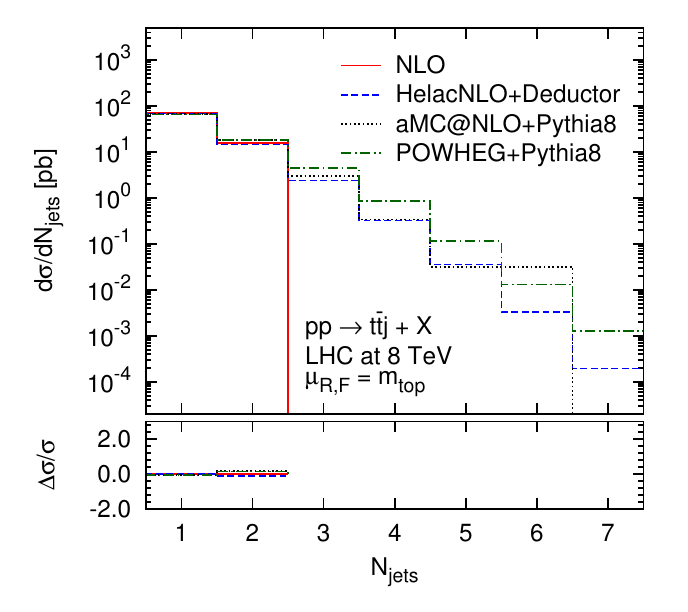}\\
\includegraphics[width=7cm]{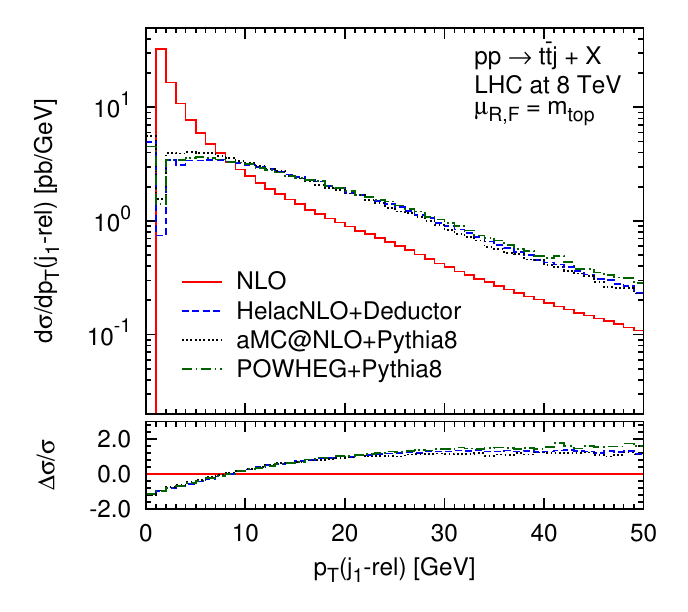}
\end{center}
\caption{Differential distributions for $t\bar{t}j$
  production. Comparison between NLO and several NLO+PS predictions.}
\label{fig:dist2}
\end{figure}

Several distributions showing a comparison between the fixed order NLO
calculation and several NLO+PS predictions 
(\textsc{aMc@NLO}+\textsc{Pythia8}~\cite{Alwall:2014hca} and
\textsc{Powheg}+\textsc{Pythia8}~\cite{Alioli:2011as}) are shown in
Figures\,\ref{fig:dist1}~and~\ref{fig:dist2}. 
$p_T(j_1-\rel)$ in Figure\,\ref{fig:dist2} is the scalar sum of the
relative transverse momenta of the particles in the first jet, defined 
with respect to the jet axis in the frame where the first jet has zero
rapidity
\begin{equation*}
p_T(j_1-\rel) = \sum_{i \in j_1} \frac{|\vec{k}_i 
\times \vec{p}(j_1)|}{|\vec{p}(j_1)|},
\end{equation*}
where $k_i$ is the momentum of the $i^{\th}$ particle in $j_1$.

We observe that distributions shown in Figure\,\ref{fig:dist1}, which are
not expected to be affected by showering effects have, to a good
approximation, the same shape as at fixed order. The resummation
effect is visible first and foremost in the transverse momentum
distribution of the $t\bar{t}j$ system and other distributions shown
in Figure\,\ref{fig:dist2}. A more detailed analysis is 
presented in \cite{in-preparation}.

\section{Conclusions}
\label{sec:conclusions} 

We have discussed recent progress towards
matching next-to-leading QCD calculations for LHC processes with
parton showers at next-to-leading logarithmic accuracy. Matched
NLO+NLL calculations will provide accurate predictions for
differential distributions and exclusive observables with experimental
cuts, and are thus essential to fully exploit the potential of the
upcoming LHC run. 

To facilitate the matching of NLO calculations with parton showers,
the subtraction terms  needed to combine virtual and real corrections
should by constructed from the splitting  functions that define the
parton shower. We have presented a subtraction scheme based  on a
parton shower with quantum interference and its implementation into
the \textsc{Helac-NLO}  software. The new subtraction scheme has been
applied to a number of  challenging processes, including the
production of up to four heavy quarks at the LHC.  The new scheme
performs well compared to established methods. It not only provides an
important internal check of multi-parton NLO calculations, but also
forms the  basis of current work on parton shower matching at NLL
accuracy.

First results of such a new NLO+NLL  calculation have been
presented. We find  that the resummation is important for a wide range
of phenomenologically relevant distributions. Work  is in progress to
further improve on the accuracy of the calculation by adding
sub-leading color effects and  spin correlations. 

\section*{Acknowledgements} 

This work was supported by the Deutsche
Forschungsgemeinschaft through the collaborative research centre
SFB-TR9 ``Computational Particle Physics",  and by the U.S. Department
of Energy under contract DE-AC02-76SF00515. We would like to thank our
collaborators Giuseppe Bevilacqua, Heribertus Bayu Hartanto, Manfred
Kraus and Michael Kubocz.  MK is grateful to SLAC and Stanford
University for their hospitality. 





\nocite{*}



\end{document}